\documentclass[12pt,a4paper]{iopart}
\usepackage{graphicx}
\expandafter\let\csname equation*\endcsname\relax
\expandafter\let\csname endequation*\endcsname\relax
\usepackage{amsmath,bm}
\usepackage{hyperref}
\usepackage[utf8]{inputenc}
\usepackage[T1]{fontenc}
\usepackage[capitalize]{cleveref}
\usepackage{xcolor}
\usepackage{breqn}

\makeatletter
\let\cref@old@eq@setnumber\eq@setnumber
\def\eq@setnumber{%
\cref@old@eq@setnumber%
\cref@constructprefix{equation}{\cref@result}%
\protected@xdef\cref@currentlabel{%
[equation][\arabic{equation}][\cref@result]\p@equation\theequation}}
\makeatother

\begin{document}

\title{Construction of Quasisymmetric Stellarators Using a Direct Coordinate Approach}

\author{R. Jorge$^{1}$, W. Sengupta$^{2}$, M. Landreman$^{1}$}
\address{$^1$Institute for Research in Electronics and Applied Physics, University of Maryland, College Park, MD 20742, USA}
\address{$^2$Courant Institute of Mathematical Sciences, New York University, New York NY 10012, USA}
\ead{rjorge@umd.edu}

\begin{abstract}
Optimized stellarator configurations and their analytical properties are obtained using a near-axis expansion approach.
Such configurations are associated with good confinement as the guiding center particle trajectories and neoclassical transport are isomorphic to those in a tokamak.
This makes them appealing as fusion reactor candidates.
Using a direct coordinate approach, where the magnetic field and flux surface functions are found explicitly in terms of the position vector at successive orders in the distance to the axis, the set of ordinary differential equations for first and second order quasisymmetry is derived.
Examples of quasi-axisymmetric shapes are constructed using a pseudospectral numerical method.
Finally, the direct coordinate approach is used to independently verify two hypotheses commonly associated with quasisymmetric magnetic fields, namely that the number of equations exceeds the number of parameters at third order in the expansion and that the near-axis expansion does not prohibit exact quasisymmetry from being achieved on a single flux surface.
\end{abstract}

\section{Introduction}
\label{sec:intro}

In order to realize sustained fusion energy, a plasma must be confined at high enough temperatures and for enough time to achieve ignition.
While several strategies have been devised to achieve this goal, in this work, we focus on toroidal devices that solely rely on the use of external magnetic fields produced by current-carrying coils to contain such high-temperature plasmas, i.e. stellarators.
One of the major challenges associated with stellarator devices is the confinement of the charged particles within the plasma.
While tokamaks rely on a continuous rotational symmetry of the magnetic field along the toroidal direction, i.e. axisymmetry, for confinement, stellarators rely on axisymmetry breaking to generate a finite rotational transform without the need for a plasma current.
In order to find a class of optimal devices that restrict charged particles to stay close to a given magnetic surface, an efficient method to explore the space of available magnetic fields is needed.

Quasisymmetric stellarators are a class of stellarator shapes with magnetic fields that can accommodate a continuous symmetry of the Lagrangian of charged particle motion without necessarily resorting to axisymmetry.
The presence of symmetries in the Lagrangian of the system, according to Noether's theorem, leads to conservation laws that constrain the trajectories of the particles, ultimately contributing to confinement in a favorable manner.
A symmetry in the Lagrangian becomes evident when the magnetic field appearing in the guiding-center Lagrangian is written in Boozer coordinates, i.e. coordinates where the magnetic field $\mathbf B$ and the streamlines of $\mathbf B \times \nabla \psi$ with $\psi$ the toroidal flux are straight \cite{Boozer1981a}.
In this case, the Lagrangian depends on position only through the poloidal or toroidal magnetic flux and the strength of the magnetic field $B$.
Thus, quasisymmetry can be stated as the condition of having $B$ depend only on some linear combination of the angles in Boozer coordinates at any flux surface \cite{Nuhrenberg1988a,Boozer1995,Garabedian1996}.
This leads to a conserved quantity analogous to the canonical angular momentum in the tokamak case that guarantees the confinement of charged particles.
Furthermore, as the drift-kinetic equation, when written in Boozer coordinates, only depends on position through flux functions and $B$, a quasisymmetric stellarator would have similar confinement properties as a tokamak when it comes to neoclassical transport \cite{Boozer1983,Helander2014}.
Additional implications of quasisymmetry include the fact that, unlike a general stellarator, the radial electric field is not determined by neoclassical transport \cite{Helander2008,Sugama2011a}, and that plasma flows (except flows as large as the ion thermal seed) larger than the ones present in a general stellarator are likely to be allowed \cite{Helander2007}.

Until recently, quasisymmetric stellarator equilibria have been found using computationally-heavy numerical optimization schemes.
An example is the Helically Symmetric Experiment (HSX), for which the design was obtained by employing non-linear optimizers with the goal of obtaining a rotational transform in the range $1 < \iota < 1.25$ free of low-order resonances, a global marginal magnetic well over most of the confinement volume, and minimal deviation from quasisymmetry along a field line out to two-thirds of the plasma boundary radius \cite{Anderson1995}.
Other quasisymmetric designs include the National Compact Stellarator eXperiment (NCSX) \cite{Zarnstorff2001} and the Chinese First Quasi-axisymmetric Stellarator (CFQS) \cite{Liu2018,Kinoshita2019}.
The difficulty of achieving exact global quasisymmetry has been discussed in \cite{Garren1991,Burby2019} where, although the existence quasisymmetric solutions other than axisymmetry has not been disproved, a set of strong constraints has been obtained.
In particular, it has been shown using an inverse near-axis approach that the relevant system of equations becomes overdetermined at third order in the expansion parameter.
However, experimental evidence from HSX has shown that even approximate quasisymmetry can still reduce the large transport at low-collisionalities typically associated with stellarators \cite{Canik2007}.

We now seek an alternative approach to obtain quasisymmetric configurations by using an expansion on the smallness of the inverse aspect ratio $\epsilon$, the so-called near-axis expansion.
While the near-axis expansion has been explored by several authors in the past \cite{Mercier1964,Solovev1970,Lortz1976}, it was only recently that such formulation was applied to the construction of quasisymmetric designs \cite{Garren1991a, Landreman2018, Landreman2019a}, with a practical procedure to obtain first and second order solutions derived in \cite{Landreman2019a,Landreman2019b}.
Notwithstanding, such constructions rely on the Garren-Boozer construction, an inverse coordinate approach that employs Boozer coordinates to determine the resulting magnetic field.
In this work, we proceed with the analytical determination of quasisymmetric magnetic in vacuum fields using a direct coordinate approach and apply it to the case of vacuum magnetic field configurations.
This involves the expansion of the magnetic flux $\psi=\psi(\mathbf r)$ in terms of the components of the position vector $\mathbf r$, while the indirect approach expands the position vector $\mathbf r=\mathbf r(\psi, \vartheta, \varphi)$ in terms of magnetic coordinates $(\psi, \vartheta, \varphi)$.
In addition to allowing for independent proofs of the properties of quasisymmetric devices, the direct method has several advantages.
Due to the fact that it does not rely on the existence of a flux surface function to define its coordinate system, the direct method allows for the construction of magnetic fields with resonant surfaces (such as magnetic islands) and can provide analytical constraints for the existence of magnetic surfaces \cite{Jorge2020}.
Also, in the direct method, the magnetic axis is defined in terms of the vacuum magnetic field, allowing the determination of a Shafranov shift and plasma beta limits when finite plasma pressure effects are included.
Finally, due to the use of an orthogonal coordinate system, explicit expressions for the magnetic field, the toroidal flux and the rotational transform can be found at arbitrary order in $\epsilon$.

Using the approach pursued here, we are able to independently confirm two hypotheses concerning quasisymmetric magnetic fields stated in Ref. \cite{Garren1991a}.
First, we show that the requirements of quasisymmetry can be satisfied up to second order in the expansion but the number of constraints overcomes the number of free functions at third and higher order.
Second, although quasisymmetry (except axisymmetry) might not be achievable globally, we show how exact quasisymmetric magnetic fields can be constructed on a particular flux surface.

In addition to the derivation of the quasisymmetry constraints and the analytical properties of quasisymmetric magnetic fields, we develop a numerical algorithm based on a pseudospectral decomposition of the flux surface components to solve the quasisymmetry equations.
Quasisymmetric magnetic field configurations, namely quasi-axisymmetric solutions are obtained at first and second order in the expansion.
In order to establish the accuracy of the obtained quasisymmetric solutions, we use the VMEC code \cite{Hirshman1983} to generate the three-dimensional equilibrium inside the computed boundary and show that the quasisymmetry-breaking modes of the magnetic field strength $B$ are small using the BOOZ\_XFORM code \cite{Sanchez2000a}.

This paper is organized as follows.
In \cref{sec:nearaxis}, the direct near-axis expansion formalism and Mercier's coordinate system are introduced.
In \cref{sec:qs} a coordinate-independent quasisymmetry constraint is derived, and Mercier's coordinates are used to evaluate the constraint at successive orders in the expansion.
In \cref{sec:qs1surf} we revisit the argument stating that, although quasisymmetry cannot be achieved globally at arbitrary order in the expansion, it can be made exact at a particular flux surface.
A formulation of quasisymmetry in terms of the quasisymmetry vector is derived in \cref{sec:qsvector} and a comparison between the formulations of quasisymmetry derived using the straight-field-line coordinates and the one used here is performed in \cref{sec:qsgbmercier}.
Finally, the numerical solution of the quasisymmetric system of equations to first and second order is the subject of \cref{sec:numqs}.
The conclusions follow.

\section{Near-Axis Expansion}
\label{sec:nearaxis}

In this section, we introduce Mercier's coordinates, a modified cylindrical coordinate system associated with the magnetic axis, and the asymptotic expansion of the magnetic field in terms of the inverse aspect ratio $\epsilon$.
The set of equations for the magnetic field and the toroidal magnetic flux is then solved up to third order in the expansion parameter $\epsilon$.
In the following, we focus on vacuum magnetic fields where $\nabla \times \mathbf B = 0$, therefore writing $\mathbf B$ in terms of a magnetic scalar potential $\phi$ as $\mathbf B = \nabla \phi$.

The magnetic axis curve is denoted as $\mathbf r_0(t)$ with $t$ a parametrization parameter, e.g. the cylindrical azimuthal angle $\Phi$.
In this section, we parametrize $\mathbf r_0$ using the arclength function $s$, where $s$ varies between $0$ and $L$ with $L$ the total length of the axis.
The transformation between $s$ and another parametrization parameter $t$ can be found using $ds/dt=|\mathbf r_0'(t)|$.
The Frenet-Ferret frame associated with the magnetic axis includes the unit tangent vector $\mathbf t=\mathbf r_0'(s)$, the unit normal vector $\mathbf n=\mathbf t'(s)/\kappa$ with $\kappa={|\mathbf r'(t)\times \mathbf r''(t)|}/{|\mathbf r'(t)|^3}$ the curvature, and the unit normal vector $\mathbf b = \mathbf t \times \mathbf n$.
We note that any curve, including the magnetic axis, can be described uniquely by the curvature $\kappa$ and torsion $\tau={\left(\mathbf r'(t) \times \mathbf r''(t)\right)\cdot \mathbf r'''(t)}/{|\mathbf r'(t) \times \mathbf r''(t)|^2}$, and that the triad $(\mathbf t, \mathbf n, \mathbf b)$ constitutes an orthonormal vector basis that obeys the Frenet-Serret equations \cite{Spivak1999}.
In the near-axis expansion using a direct coordinate approach, the radius vector $\mathbf r$ is written as \cite{Mercier1964,Solovev1970}
\begin{equation}
    \mathbf r = \mathbf r_0 + \rho(\cos \theta \mathbf n + \sin \theta \mathbf b),
\label{eq:radiusvec}
\end{equation}
with $\rho$ the distance between an arbitrary point $\mathbf r$ and $\mathbf r_0$ in the plane $(\mathbf n, \mathbf b)$ perpendicular to the axis and $\theta$ the angle measured from the normal $\mathbf n$ to $\mathbf r-\mathbf r_0$.
Although the coordinates $(\rho, \theta, s)$ are not orthogonal, an orthogonal set of coordinates $(\rho, \omega, s)$ with $\omega=\omega(\theta,s)$ can be found by defining
\begin{equation}
    \omega = \theta + \gamma(s),~\gamma(s)=\int_0^s \tau(s') ds'.
\end{equation}
In $(\rho, \omega, s)$ coordinates, the square root of the determinant of the metric tensor $\sqrt{g}$ is given by $\sqrt g= \rho h_s$ with $h_s=1-\kappa \rho \cos \theta$.

In the following, we normalize $\kappa, \tau, s$ and $\nabla$ by $R$, the minimum of the local radius of curvature of the magnetic axis, and normalize $\rho$ by $a$, the maximum perpendicular distance from the axis to the plasma boundary.
Furthermore, we introduce the expansion parameter $\epsilon$ as
\begin{equation}
    \epsilon=\frac{a}{R}\ll 1.
\end{equation}
The magnetic field $\mathbf{B}$ is normalized by the averaged field $\overline B=\int_0^L B_0(s)ds/L$ with $B_0(s)$ the strength of the magnetic field on axis, the toroidal flux $\psi$ is normalized to $\overline B R^2$ and the magnetic scalar potential is normalized to $\overline B R$.
Noting that $\mathbf{B}=\nabla \phi={\partial_\rho \phi}\mathbf e_\rho/\epsilon+{\partial_\omega \phi}\mathbf e_\omega/(\epsilon \rho)+{\partial_s \phi}\mathbf e_s/h_s$, with $\mathbf e_\rho = \cos \theta \mathbf n + \sin \theta \mathbf{b}$, $\mathbf{e}_\omega = -\sin \theta \mathbf n + \cos \theta \mathbf b$ and $\mathbf e_s = \mathbf t$, the function $\phi$ is obtained by solving Laplace's equation
\begin{equation}
    \nabla^2 \phi = \frac{1}{\epsilon^2\rho}\frac{\partial}{\partial \rho}\left(h_s \rho \frac{\partial \phi}{\partial \rho}\right)+\frac{1}{\epsilon^2\rho^2}\frac{\partial}{\partial \omega}\left(h_s\frac{\partial \phi}{\partial \omega}\right)+\frac{\partial}{\partial s}\left(\frac{1}{h_s}\frac{\partial \phi}{\partial s}\right)=0,
\label{eq:laplacephi}
\end{equation}
subject to $\mathbf B(\rho=0)=B_0(s) \mathbf e_s$.
The toroidal flux $\psi$ is found by solving $\nabla \psi \cdot \nabla \phi=0$, yielding
\begin{equation}
    \frac{\partial \phi}{\partial \rho}\frac{\partial \psi}{\partial \rho}+\frac{1}{\rho^2}\frac{\partial \phi}{\partial \omega}\frac{\partial \psi}{\partial \omega}+\frac{\epsilon^2}{h_s^2}\frac{\partial \phi}{\partial s}\frac{\partial \psi}{\partial s}=0.
\label{eq:eqpsi0phi1}
\end{equation}
Equation (\ref{eq:eqpsi0phi1}) is supplemented by the condition
\begin{equation}
    \psi = \frac{1}{L}\int (\mathbf B \cdot \nabla s) dV,
\label{eq:psinabla}
\end{equation}
with $dV$ the volume element, which sets $\psi$ equal to the toroidal flux.
We note that, up to the order considered here, the role of \cref{eq:psinabla} is to set the arbitrary multiplicative constant present in the second order expression for $\psi$ such that $\psi$ equals the toroidal flux.

We proceed by solving \cref{eq:laplacephi,eq:eqpsi0phi1} using an expansion of $\phi$ and $\psi$ in a power series of the form
\begin{align}
    \phi&=\sum_{n=0}^{\infty} \phi_{n}(\omega,s) \epsilon^n \rho^n,
\label{eq:powerser}
\end{align}
and similarly for $\psi$.
Solving \cref{eq:laplacephi} up to $O(\epsilon^3)$ we obtain
\begin{align}
    \phi(s,\omega) &\simeq \int_0^s B_0(s')ds'+\epsilon^2 \rho^2 B_0(s)\left[\phi_{20}(s)+\phi_{22}^s(s)\sin 2u + \phi_{22}^c(s) \cos 2u\right]\nonumber\\
    &+\epsilon^3 \rho^3 B_0(s)[\phi_{31}^c(s) \cos \theta + \phi_{31}^s(s) \sin \theta + \phi_{33}^c(s) \cos 3\theta + \phi_{33}^s(s) \sin 3\theta],
\label{eq:phi23}
\end{align}
with $\phi_{20}=-B_0'/4$, $u = \omega - \gamma(s)+\delta(s)$, $\phi_{22}^s=(B_0/2) u' \tanh \eta$, $\phi_{22}^c=-(B_0/4)\eta'$ and the third order components of $\phi_{31}^c$ and $\phi_{31}^s$ given by
\begin{align}
    \phi_{31}^c(s)&=-\frac{\kappa}{8}\left(\frac{\kappa'}{\kappa}+\frac{2B_0'}{B_0}+\frac{\eta'}{2}\cos 2 \delta- u' \tanh \eta \sin 2 \delta \right),
\end{align}
and
\begin{align}
    \phi_{31}^s(s)&=\frac{\kappa}{8}\left(-\tau+\frac{\eta'}{2}\sin 2 \delta+\mu u' \cos 2 \delta\right).
\end{align}
The functions $\eta(s), \delta(s), \phi_{33}^s(s)$ and $\phi_{33}^c(s)$ are the free parameters of the system that can be used to specify the shaping of the magnetic field.
We note that, at each order $n$ in $\epsilon$, additional freedom is introduced in the solution to the Laplace's equation, \cref{eq:laplacephi}, via two functions of $s$, namely $\phi_{nn}^c$ and $\phi_{nn}^s$, the coefficients of the $n$th cosine and sine Fourier series in $\theta$ of $\phi_n$, respectively.

The asymptotic series for the toroidal flux $\psi$ is derived using \cref{eq:eqpsi0phi1}, which to third order in $\epsilon$ yields
\begin{align}
    \psi &\simeq \epsilon^2 \rho^2 {\pi B_0}(\cosh \eta+\sinh \eta \cos 2 u)\nonumber\\
    &+\epsilon^3 \rho^3[\psi_{31}^c \cos \theta + \psi_{31}^s \sin \theta + \psi_{33}^c \cos 3\theta + \psi_{33}^s \sin 3\theta].
\label{eq:psi03}
\end{align}
The third order components $\Psi_{3}=B_0^{-3/2}[\psi^{c}_{31}  \psi^{s}_{31} \psi^{c}_{33} \psi^{s}_{33}]^T$ are the solution of the following set of ordinary differential equations
\begin{equation}
    \Psi^{'}_{3}=M_{1} \cos 2 \delta \Psi_{3}+\frac{B_0^{-3/2} \pi}{8}M_{2},
\label{eq:systempsi03}
\end{equation}
with $M_{1}$ the matrix
\begin{equation}
    M_{1}=
\left(
\begin{array}{cccc}
 \mu_2  & -\mu_1 +\tau  \sec 2 \delta & \frac{3 \mu_2}{2}  & -3\mu_1/2 \\
 - \mu_1 - \tau \sec 2 \delta & -\mu_2  & 3\mu_1/2 & \frac{3 \mu_2}{2}  \\
 \mu_2/2 & \mu_1/2 & 0 & 3 \tau  \sec 2 \delta  \\
 -\mu_1/2 & \mu_2/2 & -3 \tau  \sec 2 \delta  & 0 \\
\end{array}
\right)
\end{equation}
where $\mu_1 = \eta '\tan 2 \delta + 2 \mu  u'$, $\mu_2 = \eta '- 2 \mu  u' \tan 2 \delta$, $\mu=\tanh \eta$, and $M_2=[M_{21} M_{22} M_{23} M_{24}]^T$ the vector
\begin{align}
    M_{21}&=6 \kappa \cosh \eta'+\cos 2 \delta  \left[4 \sinh \eta  \left(\kappa'-12 {\phi_{33}^{c}}\right)-5 \eta ' \kappa \cosh \eta\right]\nonumber\\
    &-14 \eta ' \kappa \sinh \eta+2 \sin 2 \delta  \sinh \eta  \left[\kappa \left(5 u'-2 \tau \right)+24 {\phi_{33}^{s}}\right],\label{eq:m21}\\
    M_{22}&=\sin 2 \delta  \left[5 \eta ' \kappa \cosh \eta-4 \sinh \eta  \left(\kappa'+12 {\phi_{33}^{c}}\right)\right]\nonumber\\
    &-2 \cos 2 \delta  \sinh \eta  \left[\kappa \left(2 \tau -5 u'\right)+24 {\phi_{33}^{s}}\right]+2 \kappa \cosh \eta \left(3 \tau +2 \mu^2 u'\right),\label{eq:m22}\\
    M_{23}&=2 \cos 2 \delta  \kappa \sinh \eta'+\kappa \left[ \eta ' \sinh \eta \cos 4 \delta  -8\eta ' \cosh \eta \cos 2 \delta   \right.\nonumber\\
    &\left.+2 \sin 2 \delta  \sinh \eta  \left(\tau +8 u'\right)-2 \sin (4 \delta ) \sinh \eta  \tanh (\eta ) u'\right]-48 \cosh \eta  {\phi_{33}^{c}},\label{eq:m23}\\
    M_{24}&=-2 \sin 2 \delta  \kappa \sinh \eta'+\kappa \left[-\eta ' \sinh \eta \sin 4 \delta   +8 \sin 2 \delta  \eta ' \cosh \eta \right.\nonumber\\
    &\left.+2 \cos 2 \delta  \sinh \eta  \left(\tau +8 u'\right)-2 \cos (4 \delta ) \sinh \eta  \mu u'\right]-48 \cosh \eta  {\phi_{33}^{s}}.\label{eq:m24}
\end{align}
The four degrees of freedom present in the near-axis formalism up to $O(\epsilon^3)$ can then be used to find the scalar potential $\phi$ in \cref{eq:phi23}, the magnetic field $\mathbf{B}=\nabla \phi$ and the toroidal flux, such that $\psi$ yields a quasisymmetric stellarator shape.
The constraint in \cref{eq:psinabla} does not impose any additional constraint at this order.
Finally, we remark that, as shown in Ref. \cite{Jorge2020}, in order for a solution for $\psi$ to exist, at this order, the rotational transform on axis must not be an integer multiple of one third.

\section{Near-Axis Quasisymmetry}
\label{sec:qs}

Quasisymmetry can be defined as a magnetic field configuration for which the magnetic field strength only varies on a surface through a linear combination of Boozer angles \cite{Boozer1995}.
Although such definition can be directly employed in the case of the near-axis construction using an inverse approach based on Boozer coordinates, a formulation of quasisymmetry independent of the coordinate system used is needed for a direct coordinate approach.
To this end, we derive that a coordinate-independent formulation of the quasisymmetry constraint in \ref{app:qscond} and show that quasisymmetry is sufficient for omnigenity, i.e. that the radial guiding-center drift averages over a bounce to zero for all trapped particles in quasisymmetric fields.
This yields \cite{Helander2008}
\begin{equation}
    \frac{\mathbf B \times \nabla \psi \cdot \nabla B}{\mathbf B \cdot \nabla B}=\frac{1}{F(\psi)}.
\label{eq:qscond}
\end{equation}
We then use \cref{eq:qscond} to determine the four free functions of \cref{sec:nearaxis}, namely $\eta, \delta, \phi_{33}^c$ and $\phi_{33}^s$.


%

\subsection{Zeroth Order Quasisymmetry}

We now write the quasisymmetry condition in \cref{eq:qscond} using the third order expressions for the magnetic field $\mathbf{B}$ and the toroidal flux $\psi$ in \cref{eq:phi23,eq:systempsi03}, respectively.
Furthermore, we expand the numerator and denominator of \cref{eq:qscond} in powers of $\epsilon \rho$ as
\begin{equation}
    \frac{\sum_{n=0}^\infty \epsilon^n \rho^n \left(\mathbf B \cdot \nabla B\right)_n}{\sum_{n=1}^\infty \epsilon^n \rho^n \left(\mathbf B \times \nabla \psi \cdot \nabla B\right)_n}={\sum_{n=0}^\infty \epsilon^{n} \rho^{n} F_{n}}.
\label{eq:qscond1}
\end{equation}
Using \cref{eq:qscond1}, the lowest order quasisymmetry condition is given by $\left(\mathbf B \cdot \nabla B\right)_0=0$, which can be written explicitly as
\begin{equation}
   B_0'(s)=0.
\label{eq:qsm1}
\end{equation}
We therefore conclude that, to lowest order, the magnetic field on axis $B_0$ must be a constant in quasisymmetric magnetic fields.
A similar condition was derived using the inverse approach in \cite{Garren1991,Garren1991a}.

\subsection{First Order Quasisymmetry}
\label{sec:qs1eqs}

To next order in \cref{eq:qscond1}, we evaluate the first order component of both $\mathbf B \cdot \nabla B$ and $\mathbf B \times \nabla \psi \cdot \nabla B$, and insert them into the quasisymmetry condition of \cref{eq:qscond1}, yielding
\begin{equation}
    (\mathbf B \cdot \nabla B)_1 = F_0[(\nabla \psi \times \nabla B) \cdot  \mathbf B]_1,
\label{eq:qs00}
\end{equation}
which can be evaluated explicitly, yielding
\begin{align}
    &{ \frac{\kappa'}{\kappa} \cos \theta+ \tau [\sin \theta -\mu \sin (\theta +2 \delta)]+ \delta ' \mu \sin (\theta +2 \delta)-\frac{\eta '}{2} \cos (\theta +2 \delta)}\nonumber\\
    &=-2 \pi F_0  B_0 [\sin \theta \cosh \eta-\sin (\theta +2 \delta)\sinh \eta ],
\label{eq:qs0}
\end{align}
with $\mu = \tanh \eta$.
In order to satisfy \cref{eq:qs0} for all values of $\theta$, we equate the $\sin \theta$ and the $\cos \theta$ components of both sides of \cref{eq:qs0}, yielding
\begin{equation}
    \frac{B_0 \pi}{\kappa^2}\frac{d}{ds}\left[\kappa^2\left(\cosh \eta - \sinh \eta \cos 2 \delta\right)\right]=0,
\label{eq:qs0cond1}
\end{equation}
and
\begin{equation}
    \sin 2 \delta \tanh \eta \left(2 \pi  B_0 F_0 \cosh \eta-\delta'+\tau\right)-\frac{\kappa'}{\kappa}+\frac{\eta '}{2} \cos 2 \delta=0.
\label{eq:qs0cond2}
\end{equation}

We now seek a form of the first order quasisymmetry constraints that allows us to combine \cref{eq:qs0cond1,eq:qs0cond2} into a single equation for the variable $\sigma$, defined as
\begin{equation}
    \sigma = \sinh \eta \sin 2 \delta,
\label{eq:sigma}
\end{equation}
and the parameter $\overline \eta$ derived from \cref{eq:qs0cond1} and given by
\begin{equation}
    \overline \eta^2 = \kappa^2(\cosh \eta - \sinh \eta \cos 2 \delta).
\label{eq:eta2}
\end{equation}
As shown in \cref{sec:qsgbmercier}, the constant $\overline \eta$ reflects the magnitude by which the magnetic field strength $B$ varies on surfaces to first order in $\epsilon$.
By differentiating \cref{eq:sigma} and noting that, according to \cref{eq:eta2,eq:sigma}, $\cosh \eta$ and $\tan 2\delta$ can be written as
\begin{align}
    \cosh \eta &= \frac{\kappa^2}{2 \overline \eta^2}\left(1+\sigma^2+\frac{\overline \eta^4}{\kappa^4}\right),
\label{eq:qscosheta}
\end{align}
and
\begin{align}
    \tan 2 \delta &=\frac{2\sigma}{1+\sigma^2-\frac{\overline \eta^4}{\kappa^4}}\frac{\overline \eta^2}{\kappa^2},
\label{eq:qstan2delta}
\end{align}
respectively, we obtain the so-called sigma equation \cite{Landreman2018}
\begin{equation}
    \sigma'+2 \pi B_0 F_0\left(1+\sigma^2+ \frac{\overline \eta^4}{\kappa^4}\right)+2\tau \frac{\overline \eta^2}{\kappa^2}=0.
\label{eq:sigmaequation}
\end{equation}
The constant $F_0$ is then determined by enforcing periodic boundary conditions on $\sigma$ and an initial condition $\sigma(0)=\sigma_0$.
A derivation of \cref{eq:eta2,eq:sigma,eq:qscosheta,eq:qstan2delta} and the relation between $F_0$ and the rotational transform on axis $\iota_0$ based on a comparison with the Garren-Boozer formalism is performed in \cref{sec:qsgbmercier}, while a numerical solution of \cref{eq:sigmaequation} is carried out in \cref{sec:numqs}.

For future reference, we rewrite the magnetic scalar potential $\phi$ in \cref{eq:phi23} in terms of the newly defined quantities $\sigma$ and $\overline \eta$ taking into account the zeroth and first order quasisymmetry constraints.
This yields
\begin{align}
    \phi &\simeq B_0 s + \epsilon^2 \rho^2 \frac{B_0}{2} \left[\left(2\pi B_0 F_0 \sigma - \frac{\kappa'}{\kappa}\right)\cos2\theta-\left(2\pi B_0 F_0 \frac{\overline \eta^2}{\kappa^2}+\tau\right)\sin2\theta\right]\nonumber\\
    &+\epsilon^3\rho^3\frac{B_0 \kappa }{4}\left[\left(\pi B_0 F_0 -\frac{\kappa'}{\kappa}\right)\cos \theta-\left(\pi B_0 F_0 \frac{\overline \eta^2}{\kappa^2}+\tau\right)\sin \theta + \frac{4}{\kappa} \phi_{33}^c \cos 3 \theta+\frac{4}{\kappa}\phi_{33}^s \sin 3 \theta\right].\label{eq:phi23qs}
\end{align}
Furthermore, we note that the second order magnetic field scalar potential $\phi$ in quasisymmetric form in \cref{eq:phi23qs} is completely determined by the axis curvature, torsion and the $\sigma$ function, obtained from the solution of \cref{eq:sigmaequation}.
Therefore, by specifying an axis shape, the parameter $\overline \eta$ and the boundary condition $\sigma_0$, only two free functions ($\phi_{33}^c$ and $\phi_{33}^s$) remain to completely determine the magnetic field and the toroidal flux up to third order in $\epsilon$.

\subsection{Second Order Quasisymmetry}
\label{sec:qs2eqs}

To next order, quasisymmetry can be written as
\begin{equation}
    (\mathbf B \cdot \nabla B)_2 = F_0[(\nabla \psi \times \nabla B) \cdot  \mathbf B]_2+F_1[(\nabla \psi \times \nabla B) \cdot  \mathbf B]_1.
\label{eq:qsconst2}
\end{equation}
Equation (\ref{eq:qsconst2}) is derived by expanding the numerator and denominator in \cref{eq:qscond1} up to second order in $\epsilon$ and using \cref{eq:qsm1,eq:qs00}.
The quasisymmetry constraints are then derived by imposing that \cref{eq:qsconst2} is satisfied for all angles $\theta$.
The $\cos \theta$ and $\sin \theta$ first Fourier modes in \cref{eq:qsconst2} are multiplied by $F_1$, which yields $F_1=0$.
The three constraints corresponding Fourier coefficients of the zeroth and second Fourier modes in $\theta$ in \cref{eq:qsconst2} being zero yield the following equations
\begin{dmath}
12 {\overline \eta}^2 \kappa^6 \phi_{33}^c = -12 F_0 {\overline \eta}^2 \kappa^6 \psi_{33}^s+2 \kappa'  \kappa^6 \left({3 \tau \sigma '}-2{{\overline \eta}^2}\right)+{16 \pi ^2 B_0^2 F_0^2 {\overline \eta}^6 \kappa'}+{4 \kappa^2 {\overline \eta}^2 \left(8 \pi  B_0 F_0 {\overline \eta}^2 \kappa' \tau+3 \kappa^{'3}\right)}-{2 {\overline \eta}^2 \kappa^3 \left[5 \kappa' \left(2 \pi  B_0 F_0 \sigma \kappa'+\kappa''\right)+6 \pi  B_0 F_0 {\overline \eta}^2 \tau'\right]}+{2 {\overline \eta}^2 \kappa^4 \left[2 \pi  B_0 F_0 \sigma \kappa''+\kappa' \left(4 \tau^2-2 \pi  B_0 F_0 \left(4 \pi  B_0 F_0+3 \sigma '\right)\right)+\kappa'''\right]}+{\kappa^7 \left[4 \pi  B_0 F_0 \sigma \left({\overline \eta}^2-3 \tau \sigma '\right)+3 \tau' \sigma '\right]}-{2 \kappa^5 {\overline \eta}^2 \left[2 \pi  B_0 F_0 \left(2 \pi  B_0 F_0 \sigma \sigma '+\sigma''\right)+5 \tau \tau'\right]},
\label{eq:sandra2}
\end{dmath}
\begin{dmath}
{12 {\overline \eta}^2 \kappa^5} \phi_{33}^s = -2 \kappa^5 \left[\sigma ' \left(\kappa''-4 \pi  B_0 F_0 \sigma \kappa'\right)+2 F_0 {\overline \eta}^2 \psi_{31}^c-6 F_0 {\overline \eta}^2 \psi_{33}^c\right]-8 \pi  B_0 F_0 {\overline \eta}^4 \kappa \kappa''+24 \pi  B_0 F_0 {\overline \eta}^4 {\kappa'}^2+\kappa^6 \left[8 \pi  B_0 F_0 \sigma ' \left(\pi  B_0 F_0+\sigma '\right)-3 {\overline \eta}^2 \tau+2 \tau^2 \sigma '\right]+{\overline \eta}^2 \kappa^4 \left[2 \tau''-\pi  B_0 F_0 \left(3 {\overline \eta}^2+8 \sigma \tau'\right)\right]-4 {\overline \eta}^2 \kappa^2 {\kappa'}^2 \tau+4 {\overline \eta}^2 \kappa^3 \left(\kappa'' \tau+\kappa' \tau'\right),
\label{eq:sandra3}
\end{dmath}
and an additional constraint equation for $\psi_{31}$
\begin{dmath}
{4 F_0 {\overline \eta}^2 \kappa^6} \psi_{31}^s = 16 \pi ^2 B_0^2 F_0^2 {\overline \eta}^6 \kappa'-2 {\overline \eta}^2 \kappa^5 \left(4 \pi ^2 B_0^2 F_0^2 \sigma \sigma '+\tau \tau'\right)+4 {\overline \eta}^2 \kappa^2 \kappa' \left(4 \pi  B_0 F_0 {\overline \eta}^2 \tau+{\kappa'}^2\right)-4 {\overline \eta}^2 \kappa^3 \left[\kappa' \left(\pi  B_0 F_0 \sigma \kappa'+\kappa''\right)+\pi  B_0 F_0 {\overline \eta}^2 \tau'\right]-\kappa^6 \kappa' \left({\overline \eta}^2-2 \tau \sigma '\right)+4 {\overline \eta}^2 \kappa^4 \left[\pi  B_0 F_0 \sigma \kappa''+\kappa' \left(\pi  B_0 F_0 \sigma '+\tau^2\right)\right]+\kappa^7 \left[\pi  B_0 F_0 \sigma \left({\overline \eta}^2-4 \tau \sigma '\right)+\tau' \sigma '\right].
\label{eq:sandra1}
\end{dmath}

The systems of equations for the higher order magnetic flux, namely the third order components $\psi_{31}^c, \psi_{31}^s, \psi_{33}^c$ and $\psi_{33}^s$ present in \cref{eq:psi03}, can now be obtained in terms of lower order quantities by rewriting the system of equations $\nabla \phi \cdot \nabla \psi = 0$ using the third order quasisymmetric magnetic scalar potential $\phi$ given by \cref{eq:phi23qs,eq:sandra2,eq:sandra3}.
This yields
\begin{align}
    \Psi_3'=M_3 \Psi_3+M_4,
\label{eq:qseqs2fin}
\end{align}
where now $\Psi_3=[\psi_{31}^c \psi_{31}^s \psi_{33}^c \psi_{33}^s]$,
\begin{align}
    M_3=\left(
\begin{array}{cccc}
 \frac{2 \kappa'}{\kappa}-6 \pi  B_0 F_0 \sigma & \frac{4 \pi  B_0 F_0 {\overline \eta}^2}{\kappa^2}+3 \tau & \frac{3 \kappa'}{\kappa} & -\frac{3 \kappa^2 \sigma'}{2 {\overline \eta}^2} \\
 \frac{2 \pi  B_0 F_0 {\overline \eta}^2}{\kappa^2}-\frac{\kappa^2 \sigma'}{2 {\overline \eta}^2} & 4 \pi  B_0 F_0 \sigma-\frac{2 \kappa'}{\kappa} & \frac{3 \kappa^2 \sigma'}{2 {\overline \eta}^2} & \frac{3 \kappa'}{\kappa} \\
 \frac{\kappa'}{\kappa}-2 \pi  B_0 F_0 \sigma & -\frac{2 \pi  B_0 F_0 {\overline \eta}^2}{\kappa^2}-\tau & 0 & -\frac{3 \kappa^2 \sigma'}{2 {\overline \eta}^2} \\
 \frac{2 \pi  B_0 F_0 {\overline \eta}^2}{\kappa^2}-\frac{\kappa^2 \sigma'}{2 {\overline \eta}^2} & \frac{\kappa'}{\kappa}-2 \pi  B_0 F_0 \sigma & \frac{3 \kappa^2 \sigma }{2 {\overline \eta}^2} & 0 \\
\end{array}
\right),
\end{align}
and $M_4$ given by \cref{eq:bb1,eq:bb2,eq:bb3,eq:bb4}.
Second order quasisymmetry can then be achieved by simultaneously solving the set of five equations in \cref{eq:sandra1,eq:qseqs2fin} for the four unknowns $\psi_{31}^c, \psi_{31}^s, \psi_{33}^c$ and $\psi_{33}^s$.
The fact that the number of scalar $s$-dependent unknowns exceeds the number of $s$-dependent equations by one for second order quasisymmetry with a fixed axis shape is a phenomenon also observed using an inverse coordinate approach \cite{Garren1991a,Landreman2019b}.
In \cref{sec:numqs}, second order quasisymmetric solutions are obtained by solving \cref{eq:qseqs2fin} using a numerical optimization algorithm in order to find an axis shape that satisfies \cref{eq:sandra1}.

Extra freedom can be gained by solving for only part of the axis shape, i.e. by letting either the curvature $\kappa$ or torsion $\tau$ in \cref{eq:sandra1,eq:qseqs2fin} be a degree of freedom of the system, making the number of equations equal the number of unknowns.
However, this should be supplemented by the requirement for the magnetic axis $\mathbf r_0$ to form a smooth closed curve.
The development of a numerical tool that partly specifies the axis shape for second order quasisymmetry will be the subject of future work.

\subsection{Third and Higher Order Quasisymmetry}
\label{sec:qs3eqs}

The third order quasisymmetry constraint is derived by expanding the numerator and denominator of \cref{eq:qscond1} into its first, second and third order components, and noting that the ratio of the first and second order components obey \cref{eq:qs0,eq:qsconst2}, respectively, yielding
\begin{equation}
    (\mathbf B \cdot \nabla B)_3 - F_0[(\nabla \psi \times \nabla B) \cdot  \mathbf B]_3=F_2[(\nabla \psi \times \nabla B) \cdot  \mathbf B]_1.
\label{eq:qsconst3}
\end{equation}
As \cref{eq:qsconst3} contains first and third Fourier modes in $\theta$, by imposing that \cref{eq:qsconst3} is satisfied for all angles $\theta$ a total of four constraint equations are obtained.
However, the addition of a fourth order component to the scalar potential in \cref{eq:phi23} would only add two free functions $\phi_{44}^c$ and $\phi_{44}^s$, as the remaining components $\phi_{40}^c, \phi_{42}^c$ and $\phi_{42}^s$ are determined by Laplace's equation $\nabla^2 \phi=0$.
Therefore, at third order, a total of seven free functions, namely $\phi_{44}^c, \phi_{44}^s, \psi_{40}^c, \psi_{42}^c, \psi_{42}^s, \psi_{44}^c, \psi_{44}^s$, and nine equations, namely four from \cref{eq:qsconst3} and five from $\nabla \phi \cdot \nabla \psi = 0$, are present, yielding an overdetermined system of equations.

A similar argument can be made to show that higher order quasisymmetric fields always yield an overdetermined system of equations.
While the quasisymmetry constraint in \cref{eq:qscond} results in $n+1$ equations at each order $n$, there is only freedom to specify two free functions at each successive order in $\epsilon$.
Therefore, even if the axis shape is let completely unspecified, an overdetermined system is always found at $n>2$.

\section{Quasisymmetry on a Flux Surface}
\label{sec:qs1surf}

We now show that, although quasisymmetry yields an overdetermined system at higher orders in the expansion, it might be still attainable exactly at one flux surface.
While this statement was first shown in Ref. \cite{Garren1991a} using an inverse coordinate approach, in this section, we use a direct coordinate approach to independently show how to construct quasisymmetric flux surfaces.

We first choose a particular flux surface, i.e. we pick a value for $\psi(\rho, \omega, s)=\psi_0$.
This allows to obtain an expression for $\rho=\rho(\psi_0,\omega,s)$ and rewrite quasisymmetry condition in \cref{eq:qscond} in terms of two variables only, namely $\omega$ and $s$.
Such expression is then Fourier expanded in $\theta$, yielding
\begin{align}
    &\sum_{j=0}^\infty \left[(\mathbf B \cdot \nabla B)_{j}^c-F(\mathbf B \times \nabla \psi \cdot \nabla B)_{j}^c\right]\cos (j \theta)\nonumber\\
    &+\left[(\mathbf B \cdot \nabla B)_{j}^s-F(\mathbf B \times \nabla \psi \cdot \nabla B)_{j}^s\right]\sin(j \theta)=0,
\label{eq:qscondMercier}
\end{align}
where $F$ is a constant, $(\mathbf B \cdot \nabla B)_{j}^c$ and $(\mathbf B \cdot \nabla B)_{j}^s$ are the $j$th coefficients (functions of $s$) of the cosine and sine Fourier series of $\mathbf B \cdot \nabla B$, respectively, and $(\mathbf B \times \nabla \psi \cdot \nabla B)_{j}^c$ and $(\mathbf B \times \nabla \psi \cdot \nabla B)_{j}^s$ are the $j$th coefficients (functions of $s$) of the cosine and sine Fourier series of $\mathbf B \times \nabla \psi \cdot \nabla B$, respectively.
At each Fourier mode $j$, \cref{eq:qscondMercier} shows that only two constraints need to be satisfied, namely $(\mathbf B \cdot \nabla B)_{j}^c=F(\mathbf B \times \nabla \psi \cdot \nabla B)_{j}^c$ and $(\mathbf B \cdot \nabla B)_{j}^s=F(\mathbf B \times \nabla \psi \cdot \nabla B)_{j}^s$, reducing the number of constraints to be satisfied at each order in $j$ from $j+1$ to $2$.
As seen in \cref{sec:qs3eqs}, at each order $j$ in $\epsilon$, there is enough freedom to specify two free functions, namely $\phi_{jj}^c$ and $\phi_{jj}^s$.
Therefore, the two quasisymmetry constraints on a particular flux-surface can, in principle, be satisfied by the two free functions $\phi_{jj}^c$ and $\phi_{jj}^s$, showing that the near-axis expansion does not prohibit the existence of quasisymmetry on a particular flux surface.

\section{Quasisymmetry Vector}
\label{sec:qsvector}

In \cref{sec:qs} the quasisymmetry constraint, \cref{eq:qscond}, was found using the equations of motion for the guiding-center.
In this section, however, we explore an equivalent formulation uniquely in terms of a continuous symmetry $\mathbf u$ of the guiding center Lagrangian and its associated conservation properties.
For a more detailed analysis of continuous symmetries of Hamiltonian systems and, in particular, the guiding-center Hamiltonian, see Refs. \cite{Arnold1978,Burby2019}, respectively.

When referring to symmetries in the system, we imply a change of coordinates that leaves the guiding-center Lagrangian $\mathcal{L}$ independent of one coordinate, thus yielding the conservation of a certain quantity according to Noether's theorem.
More precisely, there is a local conserved quantity if the Lie derivative of both the Hamiltonian $H$ and its symplectic form $\omega$ along a direction $\mathbf u$ vanishes, i.e. $L_{\mathbf u}H=L_{\omega}H=0$.
As shown in Ref. \cite{Burby2019}, a symmetry of the guiding-centre Lagrangian $\mathcal{L}$ of \cref{eq:gclag}, i.e. quasisymmetry, requires the existence of a nonzero vector $\mathbf u$, hereby called the quasisymmetry vector, satisfying the following conditions
\begin{align}
    \mathbf{u} \cdot \nabla B &= 0,\label{eq:qsu1}\\
    \nabla \times (\mathbf u \times \mathbf B) &= 0,\label{eq:qsu2}\\
    (\nabla \times \mathbf{B})\times \mathbf u+\nabla (\mathbf u \cdot \mathbf B)&=0.\label{eq:qsu3}
\end{align}
We note that the condition in \cref{eq:qsu1} can be replaced by
\begin{equation}
    \nabla \cdot \mathbf u = 0.
\label{eq:qsu4}
\end{equation}
The condition in \cref{eq:qsu3} can be further simplified by noting that, in vacuum, $\nabla \times \mathbf B = 0$, yielding
\begin{equation}
    \mathbf u \cdot \mathbf B = C,
\label{eq:qsu31}
\end{equation}
with $C$ a freely chosen constant.
The fact that $C$ can be chosen arbitrarily is due to the fact that the vector $\mathbf u$ can be freely rescaled while still satisfying \cref{eq:qsu1,eq:qsu2,eq:qsu3,eq:qsu4}.
Furthermore, we note that \cref{eq:qsu2} can be written in terms of a scalar potential $\Gamma$ as
\begin{equation}
    \mathbf B \times \mathbf u = \nabla \Gamma.
\label{eq:qsu21}
\end{equation}
Taking the scalar product of \cref{eq:qsu21} with $\mathbf B$ we find that $\mathbf B \cdot \nabla \Gamma = 0$, which allows us to set $\Gamma=\Gamma(\psi)$ (for more details see Ref. \cite{Burby2019}).
Using \cref{eq:qsu31,eq:qsu21}, the magnetic field $\mathbf B$ can be written in terms of $\mathbf u$ and $\psi$ as
\begin{equation}
    \mathbf B = C\frac{\mathbf u}{|\mathbf u|^2}+\Gamma'(\psi)\frac{\mathbf u \times \nabla \psi}{|\mathbf u|^2}.
\end{equation}
Conversely, we write the quasisymmetry vector $\mathbf u$ in terms of $\mathbf B$ and $\psi$ by crossing \cref{eq:qsu21} with $\mathbf B$, yielding
\begin{equation}
    \mathbf u = C\frac{\mathbf B}{B^2}-\Gamma'(\psi)\frac{\mathbf B \times \nabla \psi}{B^2},
\label{eq:qsuu1}
\end{equation}
The first term in \cref{eq:qsuu1} describes the path of the magnetic field line, while the second term in \cref{eq:qsuu1} changes the direction of the vector $C \mathbf B/B^2$ so that the line describing the vector $\mathbf u$ on a given surface of constant $\psi$ becomes closed.
The fact that the lines of $\mathbf u$ or, equivalently, the direction of quasisymmetry in the torus, are closed can be derived by noting that the flux along the poloidal Boozer direction of the vector $\mathbf u$ is zero \cite{Isaev1994}.

An equivalent expression for $\mathbf u$ can be obtained by crossing \cref{eq:qsu21} with $\nabla B$ and using \cref{eq:qsu1}, yielding
\begin{equation}
    \mathbf u=\Gamma'(\psi)\frac{\nabla \psi \times \nabla B}{\mathbf B \cdot \nabla B}.
\label{eq:qsuu}
\end{equation}
The quasisymmetry condition in \cref{eq:qscond} can then be obtained by taking the scalar product of \cref{eq:qsuu1} with $\nabla B$, use \cref{eq:qsu1}, and identify $\Gamma'(\psi)=C F(\psi)$ or, equivalently, the scalar product of \cref{eq:qsuu} with $\mathbf B$ and use \cref{eq:qsu21}.
We also derive from \cref{eq:qsuu1,eq:qsuu} the condition
\begin{equation}
    \mathbf u \cdot \nabla \psi = 0,
\end{equation}
showing that, as expected, the field lines of $\mathbf u$ lie on surfaces of constant $\psi$.

In order to derive the form of the quasisymmetry vector $\mathbf u$ near the axis, we calculate $\mathbf u$ using \cref{eq:qsuu1} and use the constraint in \cref{eq:qsu1}.
To lowest order in $\epsilon$, in Mercier's coordinates, \cref{eq:qsuu1} yields
\begin{equation}
    \mathbf{u}=\frac{C}{B_0}\mathbf e_s+O(\epsilon).
\end{equation}
Noting that $\nabla B \simeq B_0'(s) \mathbf e_s+B_0 \kappa(\cos \theta \mathbf e_\rho -\sin \theta \mathbf e_\omega)+O(\epsilon)$, from \cref{eq:qsu1} we find that, to lowest order,
\begin{equation}
    B_0'(s)=0,
\end{equation}
showing that the magnetic field on axis should be constant, consistent with \cref{eq:qsm1}.

To first order in $\epsilon$, we obtain
\begin{align}
    \mathbf u &\simeq \frac{C h_s}{B_0}\mathbf e_s - \frac{\epsilon \rho}{2B_0}\left(2 \tanh \eta \sin 2u \left(2 \pi  \Gamma' B_0 \cosh \eta-C u' \right)+C \eta' \cos 2u\right)\mathbf{e}_\rho \nonumber\\
    &+ \frac{\epsilon \rho}{2B_0}\left[-4 \pi  \Gamma' B_0 (\sinh \eta \cos 2u+\cosh \eta)+2 C u' \tanh \eta  \cos 2u+C \eta' \sin 2u\right]\mathbf e_\omega.
\label{eq:firstordu}
\end{align}
Imposing $\mathbf u \cdot \nabla B = 0$ with $\mathbf u$ given by \cref{eq:firstordu} yields
\begin{align}
    &\kappa C \cos \theta\left[\frac{\kappa'}{\kappa}- \sin 2\delta \sinh \eta \left( \frac{2 \pi  \Gamma'B_0}{C} - \frac{u'}{\cosh \eta}\right)- \cos 2\delta \frac{\eta'}{2}\right]\nonumber\\
    &+\kappa C^2 \sin \theta \left[ \frac{2 \pi  \Gamma' B_0}{C} (\cosh \eta-\cos 2\delta \sinh \eta)+\tau+\frac{\eta'}{2}\sin 2\delta + u' \cos 2\delta \tanh \eta\right]=0.
\label{eq:qsconst1u}
\end{align}
The first order quasisymmetry constraint is then obtained by imposing that both the coefficients of $\cos \theta$ and $\sin \theta$ in \cref{eq:qsconst1u} vanish in order to satisfy \cref{eq:qsconst1u} for all values of $\theta$.
By multiplying the $\cos \theta$ coefficient by $\cosh \eta-\sinh \eta \cos 2 \delta$ and adding the $\sin \theta$ coefficient multiplied by $\sin 2 \delta \sinh \eta$ we recover the first order quasisymmetry constraint in \cref{eq:qs0cond1}.
The second constraint in \cref{eq:qs0cond2} is recovered from the vanishing of the $\cos \theta$ coefficient in \cref{eq:qsconst1u} and identifying the zeroth order $\Gamma'$ as $\Gamma'=C F_0$.
Finally, we simplify the quasisymmetry vector $\mathbf u$ using the quasisymmetry constraints in \cref{eq:qs0cond1,eq:qs0cond2} to rewrite $\delta'$ and $\eta'$ in terms of $\delta$ and $\eta$, yielding
\begin{align}
    \mathbf u \simeq &-\frac{C \rho}{B_0} \left[\sin 2\theta \left(2 \pi  B_0 F_0 \cosh \eta+\tau\right)+\cos 2\theta\frac{\kappa'}{\kappa} \right]\mathbf e_\rho\nonumber\\
    &-\frac{C \rho}{B_0}  \left[\cos 2\theta \left(2 \pi  B_0 F_0 \cosh \eta+\tau\right)-\sin 2\theta \frac{\kappa'}{\kappa}+2 \pi  B_0 F_0 \cosh \eta\right]\mathbf e_\omega \nonumber\\
    &+\frac{C}{B_0}(1-\epsilon \rho \kappa \cos \theta)\mathbf e_s.
\label{eq:nearaxisu}
\end{align}
The second order quasisymmetry vector and the associated quasisymmetry constraints can be found in a similar way.

\section{Comparison With the Garren-Boozer Construction}
\label{sec:qsgbmercier}

In this section, we compare the quasisymmetric near-axis framework developed in the previous sections using the direct method, with an inverse coordinate formulation using straight-field-line coordinates, namely Boozer coordinates $(\psi,\vartheta,\varphi)$.
While only first order quasisymmetry comparisons are shown, the methods developed here are applicable to higher order formulations.
We note that the equivalence between the first order Garren-Boozer construction and the direct coordinate approach for the general case (without quasisymmetry) has already been established in Ref. \cite{Jorge2020}.

Up to first order in $\epsilon$, in the inverse coordinate approach, a quasisymmetric magnetic field can be described by the position vector \cite{Landreman2019b}
\begin{equation}
    \mathbf r = \mathbf r_0 + \sqrt{\frac{\psi}{\pi B_0}}\left[\frac{\overline \eta}{\kappa}\cos \overline \vartheta \mathbf n + \frac{\kappa}{\overline \eta}\left(\sin \overline \vartheta+\sigma \cos \overline \vartheta\right)\mathbf b\right]+O(\epsilon^2),
\label{eq:rqsboozer}
\end{equation}
with $\overline \vartheta = \vartheta - N \varphi$.
By equating the position vectors in the direct and inverse approaches, namely \cref{eq:radiusvec,eq:rqsboozer}, and using the lowest order expression for $\psi$ in \cref{eq:psi03}, we obtain
\begin{equation}
    \cos \overline \vartheta = \frac{\kappa}{\overline \eta}\frac{\cos \theta}{\sqrt{\cosh \eta+\sinh \eta \cos(2\theta+2\delta)}},
\label{eq:cosvarthetaqs}
\end{equation}
and
\begin{equation}
    \sin \overline \vartheta = \frac{\overline \eta}{\kappa}\frac{\sin \eta}{\sqrt{\cosh \eta+\sinh \eta \cos(2\theta+2\delta)}}-\sigma \cos \vartheta.
\label{eq:sinvarthetaqs}
\end{equation}
By squaring and adding \cref{eq:cosvarthetaqs,eq:sinvarthetaqs}, we find
\begin{align}
    &\cos 2 \theta \left[\frac{\kappa^2}{2\overline \eta^2}\left(1+\sigma^2-\frac{\overline \eta^4}{\kappa^3}\right)-\sinh \eta \cos2\delta\right]+\sin 2 \theta \left(\sinh \eta \sin 2 \delta - \sigma\right)\nonumber\\
    &+\frac{\kappa^2}{\overline \eta^2}\left(1+\sigma^2+\frac{\overline \eta^4}{\kappa^4}\right)-\cosh \eta = 0.
\label{eq:equalityqs1gb}
\end{align}
The transformation between first order quasisymmetry variables $(\eta,\delta)$ in the direct coordinate approach to the variables $(\sigma, \overline \eta)$ in the inverse approach can be found by integrating \cref{eq:equalityqs1gb} over $\theta$ with the multiplication weights $1, \cos 2\theta$ and $\sin 2 \theta$.
Using this method, we retrieve the system of equations in \cref{eq:eta2,eq:sigma,eq:qscosheta,eq:qstan2delta} used to obtain the sigma equation in \cref{eq:sigma}.

The sigma equation derived using an inverse coordinate approach is given by \cite{Landreman2019b}
\begin{equation}
    \frac{d \sigma}{d \varphi}+(\iota_0-N)\left(\frac{\overline \eta^4}{\kappa^4}+1+\sigma^2\right)+2 \frac{G_0}{B_0}\frac{\overline \eta^2}{\kappa^2}\tau = 0,
\label{eq:sigmab1}
\end{equation}
with $G_0$ is the solution of $s'(\varphi)=G_0/B_0$ or, equivalently,
\begin{equation}
    G_0 = \frac{B_0 L}{2\pi},
\label{eq:G0const}
\end{equation}
and $L=\int_0^{2\pi} s'(\phi) d\phi$ the total length of the magnetic axis.
Due to the form of the vacuum magnetic field in Boozer coordinates $\mathbf B = G_0 \nabla \varphi$, the constant $G_0$ can also be defined as the net poloidal coil current through the hole defined by a toroidal flux surface multiplied by $\mu_0/2\pi$.
Using the chain rule to rewrite the derivative in \cref{eq:sigmab1} from $\varphi$ to $s$ yields
\begin{equation}
    \frac{d \sigma}{d s}+\frac{2 \pi}{L}(\iota_0-N)\left(\frac{\overline \eta^4}{\kappa^4}+1+\sigma^2\right)+2\tau\frac{\overline \eta^2}{\kappa^2} = 0,
\label{eq:sigmab2}
\end{equation}
Comparing \cref{eq:sigmab2} with the $\sigma$ equation in \cref{eq:sigmaequation}, we find the relation between $\iota_0$ and $F_0$
\begin{equation}
    F_0 = \frac{\iota_0-N}{B_0 L}.
\label{eq:f0i0rel}
\end{equation}

Having established \cref{eq:f0i0rel}, we now derive the first order quasisymmetry constraints using the transformation between Mercier $(\rho, \omega, s)$ and Boozer $(\psi, \vartheta, \varphi)$ coordinates by requiring that the magnetic field strength only varies on a flux surface via a linear combination of Boozer angles, namely
\begin{equation}
    B=B(\psi,M\vartheta-N \varphi),
\label{eq:qsboozer1}
\end{equation}
with $M$ and $N$ integers.
As discussed in \cite{Landreman2018}, the integer $N$ can be determined directly from the axis shape by counting the number of times that the normal vector $\mathbf{n}$ rotates poloidally around the magnetic axis after one toroidal transit along the axis.
Furthermore, due to the fact that $\nabla_\perp B = \kappa B \mathbf n$ and $\kappa$ must be nonzero for some of its length, only $M=1$ quasisymmetry is allowed near the axis.

The magnetic field strength $B=|\nabla \phi|$ in Mercier coordinates, up to first order in $\epsilon$, is given by
\begin{equation}
    B=B_0(s)\left(1- \kappa \epsilon \rho \cos \theta\right).
\label{eq:bstrength1}
\end{equation}
The function $\rho$ in \cref{eq:bstrength1} can be put in terms of $(\psi, \theta, s)$ using the transformation between Mercier and Boozer coordinates \cite{Jorge2020}
\begin{align}
    \psi(\rho,\omega,s) &= {B_0 \pi \epsilon^2\rho^2}\left(e^{\eta} \cos^2 u+e^{-\eta}\sin^2 u\right)+O(\epsilon^3)\label{eq:psiboozer}\\
    \varphi(\rho,\omega,s) &= \frac{1}{G_0}\int_0^s B_0(s') ds'+O(\epsilon),\label{eq:phiboozer}\\
    \vartheta(\rho,\omega,s) &= \arctan\left(e^{-\eta}\tan u\right)- v(s)+\frac{\iota_0}{G_0}\int_0^s B_0(s') ds'+O(\epsilon).\label{eq:thetaboozer}
\end{align}
with the rotational transform on axis $\iota_0$
\begin{equation}
    \iota_0=\frac{v(L)-\left[\delta(L)-\delta(0)\right]}{2\pi}+N,
\label{eq:iota0eq}
\end{equation}
and $v(s)$ the function
\begin{equation}
    v(s)=\int_0^s \frac{\delta'(x)-\tau(x)}{\cosh \eta(x)}dx.
\label{eq:defvs}
\end{equation}
Furthermore, the $u$ dependent terms in $\rho$ can be written in terms of the Boozer toroidal angle $\varphi$ using the following trigonometric identities, valid for $|u|<\pi/2$
\begin{equation}
    \cos\left[\arctan(e^{-\eta} \tan u)\right]=\frac{e^{\eta/2}\cos u}{\sqrt{e^{\eta}\cos^2 u + e^{-\eta}\sin^2 u}},
\end{equation}
and
\begin{equation}
    \sin\left[\arctan(e^{-\eta} \tan u)\right]=\frac{e^{-\eta/2}\sin u}{\sqrt{e^{\eta}\cos^2 u + e^{-\eta}\sin^2 u}}.
\end{equation}
This yields the following form for the magnetic field strength
\begin{align}
    B&=B_0(s)\left(1- \sqrt{\frac{\psi}{\pi B_0}}\left[c_1(s)\cos(\overline \vartheta)+c_2(s)\sin(\overline \vartheta)\right]\right),
\label{eq:magfieldboozmerc}
\end{align}
where the coefficients $c_1$ and $c_2$ can be written as
\begin{align}
    c_1=\kappa(s)\left(e^{-\eta/2}\cos \delta \cos f+e^{\eta/2}\sin \delta \sin f\right),
\label{eq:a1bfield}
\end{align}
and
\begin{align}
    c_2=\kappa(s)\left(e^{\eta/2}\sin \delta \cos f-e^{-\eta/2}\cos \delta \sin f\right),
\label{eq:a2bfield}
\end{align}
and $f$ is the angle
\begin{equation}
    f(s)=v(s)-[v(L)-\delta(L)]\frac{s}{L},
\end{equation}
where \cref{eq:G0const} was used to simplify the expression for $\varphi$ as $\varphi=2 \pi s/L$.
Noting that, due to \cref{eq:phiboozer}, the arclength $s$ is a function of toroidal Boozer angle $\varphi$ only, we can conclude that the three quantities $B_0, c_1$ and $c_2$ should be constant in order to satisfy \cref{eq:qsboozer1}.
As $\delta(0)=v(0)=0$, we conclude that $c_2=0$, yielding the first quasisymmetry constraint
\begin{equation}
    \tan [f(s)] = e^{\eta(s)} \tan [\delta(s)].
\label{eq:boozerconstqs1}
\end{equation}
By deriving \cref{eq:boozerconstqs1} with respect to $s$, we recover the first quasisymmetry constraint in the form of the $\sin \theta$ coefficient of \cref{eq:qsconst1u} by identifying $[v(L)-\delta(L)+\delta(0)]/L=2\pi \Gamma' B_0/C$ which, using the fact that $\Gamma' = C F_0$ and $F_0=2\pi(\iota_0-N)/(L B_0)$ yields the definition of rotational transform on axis in \cref{eq:iota0eq}.
The second quasisymmetry constraint is derived by imposing $c_1$ to be a constant.
By squaring \cref{eq:a1bfield} and using \cref{eq:boozerconstqs1}, we obtain
\begin{equation}
    c_1^2={\overline \eta}^2=\kappa^2\left(\cosh \eta - \sinh \eta \cos 2 \delta\right),
\label{eq:boozerconstqs2}
\end{equation}
retrieving the second quasisymmetry constraint in Mercier's coordinates from \cref{eq:qs0cond1}.
Similarly, the quasisymmetry constraint in \cref{eq:qs0cond2} can be derived by taking the derivative of \cref{eq:boozerconstqs1} with respect to the arclength and using \cref{eq:boozerconstqs2}, recovering the relation in \cref{eq:f0i0rel} between $F_0$ in \cref{eq:qs0cond2} and $\iota_0$ in \cref{eq:iota0eq}.

As an aside, we note that \cref{eq:boozerconstqs2} allows us to derive a first order approximation to the shape parameter $\eta$ when the curvature variations along the axis are small, i.e. the expression for $\tilde \eta=\eta-\eta_0\ll \eta_0$ when $\tilde \kappa = \kappa - \kappa_0 \ll \kappa_0$ to first order in $\tilde \eta/\eta_0 \sim \tilde \kappa/\kappa_0$.
This yields
\begin{equation}
    \tilde \eta = \frac{2}{\mu_0}\frac{\tilde \kappa}{\kappa_0},
\end{equation}
with $\mu_0 = \tanh \eta_0$.
The zeroth order solution for $\eta_0$ is given by
\begin{equation}
    \eta_0 = -2 \ln\left(\frac{\overline \eta}{\kappa_0}\right).
\end{equation}

Finally, we take advantage of the Garren-Boozer construction in order to determine the near-axis quasisymmetry vector in Boozer coordinates and compare it with \cref{eq:nearaxisu}.
As the quasisymmetry vector is the vector that leaves the guiding-center Lagrangian $\mathcal L$ invariant, i.e.
\begin{equation}
    \mathbf u \cdot \nabla \mathcal L = 0,
\end{equation}
and a symmetry is present in the Lagrangian when it only depends on the Boozer angles as $\chi=M \vartheta - N \varphi$, we can write
\begin{equation}
    \mathbf u = D\left(\frac{\partial \mathbf r}{\partial \varphi}\right)_{\chi},
\label{eq:uqsbooz}
\end{equation}
such that $\mathbf u \cdot \nabla = D \partial/\partial \varphi$ with $D$ a constant.
We note that the partial derivative in \cref{eq:uqsbooz} is performed at constant $\chi$ and $\mathbf r$ is the position vector, which to first order in $\epsilon$ in the Garren-Boozer construction is given by \cref{eq:rqsboozer}.
By expressing $\mathbf B$ in terms of Boozer coordinates, it is possible to show that the quasisymmetry vector in \cref{eq:uqsbooz} satisfies the quasisymmetry constraints in \cref{eq:qsu1,eq:qsu2,eq:qsu3}.
Taking $\mathbf r$ to be given by \cref{eq:rqsboozer}, we obtain the first order quasisymmetry vector in Boozer coordinates
\begin{equation}
    \mathbf u = D\frac{L}{2\pi}\mathbf e_s+O(\epsilon),
\end{equation}
where we used the fact that $s'(\varphi)=G_0/B_0=L/2\pi$.
We therefore identify the constant $C$ in \cref{eq:qsu31} as
\begin{equation}
    C = D G_0 = D \frac{B_0 L}{2\pi}.
\end{equation}

\section{Numerical Quasisymmetric Solutions}
\label{sec:numqs}

We now show how to numerically solve the first and second order quasisymmetry equations for the toroidal flux $\psi$ derived in \cref{sec:qs1eqs,sec:qs2eqs}, respectively, using a pseudospectral decomposition similar to the one used in Ref. \cite{Landreman2019a}.
The construction of quasisymmetric shapes is performed by adding a quasisymmetry module to the Stellarator Equilibrium Near Axis Code (SENAC) \cite{SENAC}.
For a description of SENAC and a comparison with the VMEC equilibrium code, see Ref. \cite{Jorge2020}.

\subsection{Numerical Method}

For a practical solution of the quasisymmetry system of equations, we separately solve the first and second order system, as the first order solution for $\sigma$ is obtained using the sigma equation in \cref{eq:sigmaequation}, which is decoupled from the second order solution for $\psi_3$ in \cref{eq:sandra1,eq:qseqs2fin}.
At first order, we consider the inputs to be $\overline \eta$, $\sigma(s=0)=\sigma_0$ and the shape of the magnetic axis.
By only considering stellarator symmetric configurations, we set $\sigma_0 = 0$.
The outputs are $\sigma$ and $2 \pi B_0 F_0$ or, equivalently, the rotational transform on axis $\iota_0$.
At second order, there are no additional inputs and there are four additional outputs, namely $\psi_{31}^c, \psi_{31}^s, \psi_{33}^c$ and $\psi_{33}^s$.

The axis curve $\mathbf r_0$ is parametrized by the angle $\Phi_a$, where $0 \le \Phi_a < 2\pi$, and is described using a cylindrical coordinate system
\begin{equation}
    \mathbf r_0 = R(\Phi_a) \mathbf e_R(\Phi_a)+Z(\Phi_a)\mathbf e_Z,
\label{eq:r0param}
\end{equation}
with $(\mathbf e_R, \mathbf e_\Phi, \mathbf e_Z)$ the cylindrical unit basis vectors.
Assuming stellarator symmetry and a closed axis curve, we write the radial and vertical components of $\mathbf r_0$ as
\begin{equation}
    R=\sum_{j}R_j \cos(j N_{fp} \Phi_a),~Z=\sum_{j}Z_j \sin(j N_{fp} \Phi_a),
\label{eq:rzaxis}
\end{equation}
where $N_{fp}$ is an integer representing the number of fields periods of the axis.
The curvature $\kappa$ and torsion $\tau$ are then computed using \cref{eq:r0param}.
The construction of a constant $\psi$ surface is performed using the SENAC code, which is able to generate an appropriate VMEC input file.
In this case, a cylindrical system is employed for the position vector $\mathbf r$, namely
\begin{equation}
    \mathbf r = R \mathbf e_R(\Phi) + Z \mathbf e_Z(\Phi),
\end{equation}
where now $(R, \Phi, Z)$ are the standard cylindrical coordinates.
The transformation between cylindrical, Mercier and VMEC coordinates and its implementation in SENAC is described in Ref. \cite{Jorge2020}.
The difference between the ideal solution at some order in $\epsilon$ and the resulting equilibrium obtained from VMEC when plugging a small but finite value for $\psi$ is not taken into account in this work.
As noted in \cite{Landreman2019b}, this might result on the appearance of finite mirror modes on the axis and other artifacts that need to be taken into account when generating a finite-minor-radius boundary.
A systematic method to examine this effect in the direct coordinate approach is left for future work.

At first order, we rewrite the derivatives with respect to the arclength in \cref{eq:sigmaequation} in terms of derivatives with respect to $\Phi_a$ in order to directly use the expressions for the curvature and torsion in terms of the angle $\Phi_a$.
Equation (\ref{eq:sigmaequation}) is then solved using Newton iteration with a pseudo-spectral collocation discretization.
To this end, we discretize $\sigma, \kappa$ and $\tau$ into a uniform grid of $N_{\Phi_a}$ points, $\Phi_{a_j}=(j-1)2\pi/N_{\Phi_a}$ where $j=1 \dots N_{\Phi_a}$.
The vector of unknowns has a total of $N_{\Phi_a}+1$ elements and is given by $[\iota, \sigma_1, ..., \sigma_{N_{\Phi_a}}]^T$ with $\sigma_j=\sigma(\Phi_{a_j})$.
The system of $N_{\Phi_a}+1$ equations is then given by \cref{eq:sigmaequation} at each angle $\Phi_{a_j}$ and by requiring $\sigma_1=\sigma(0)$ which, assuming stellarator symmetry, yields $\sigma(0)=0$.
The derivative of $\sigma$ with respect to $\Phi_a$ is discretized using the Fourier pseudo-spectral differentiation matrix \cite{Weideman2000}, and the system is solved with Newton's method.

A similar approach is followed for the solution of the second order quasisymmetry equations in \cref{eq:qseqs2fin}.
The derivatives are rewritten in terms of the angle $\Phi_a$ while the functions $\psi_{31}^c, \psi_{31}^s, \psi_{33}^c$ and $\psi_{33}^s$ are discretized on the same uniform grid of $N_{\Phi_a}$ points as $\sigma, \kappa$ and $\tau$.
As in the first order solution, the axis shape in \cref{eq:r0param,eq:rzaxis} is taken to be an input.
This yields an overdetermined system of equations, as there is an additional constraint for second order quasisymmetry, namely \cref{eq:sandra1}.
In order to satisfy \cref{eq:sandra1} as well as solving for $\psi_{31}^c, \psi_{31}^s, \psi_{33}^c$ and $\psi_{33}^s$ using \cref{eq:qseqs2fin}, we carry out a numerical optimization procedure using the least squares error in \cref{eq:sandra1} as the objective function.
Defining $\psi_{31}^{sQS}$ as the expression for $\psi_{31}^s$ in \cref{eq:sandra1}, the discretized version of the objective function $f_O$ used for the minimization procedure is then given by
\begin{equation}
    f_O=\sum_{j=1}^{N_{\Phi_a}}(\psi_{31j}^{sQS}-\psi_{31j}^s)^2.
\label{eq:objf0}
\end{equation}
The axis shape and the value of $\overline \eta$ are then varied in order to minimize the objective function $f_O$ in \cref{eq:objf0}.
Finally, we note that for both the first order and second order calculations, a total of $N_{\Phi_a}=200$ points are used.

\subsection{Numerical Quasisymmetric Solutions}

\begin{figure}
    \centering
    \includegraphics[trim=130 30 70 60,clip,width=.6\textwidth]{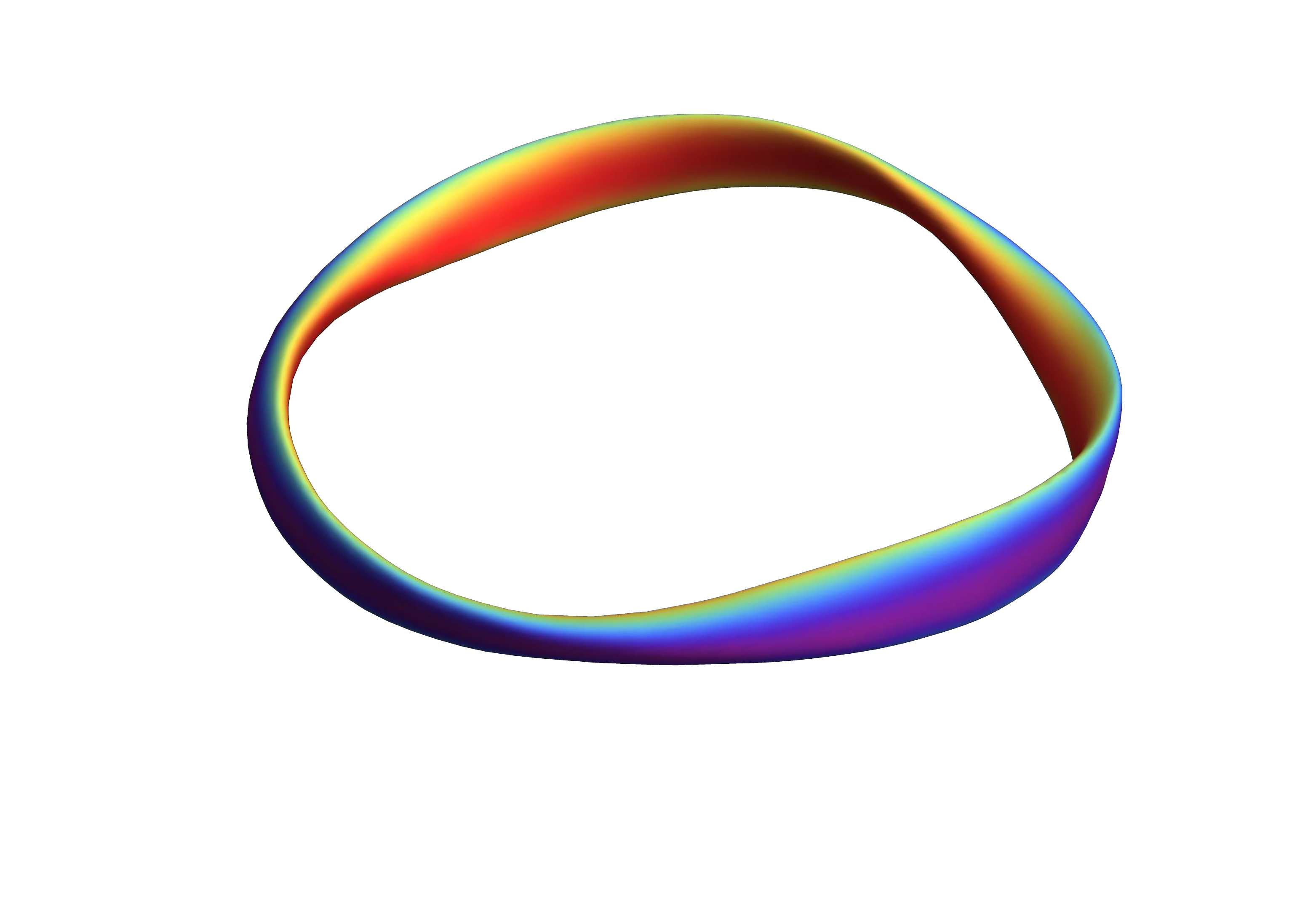}
    \includegraphics[width=.34\textwidth]{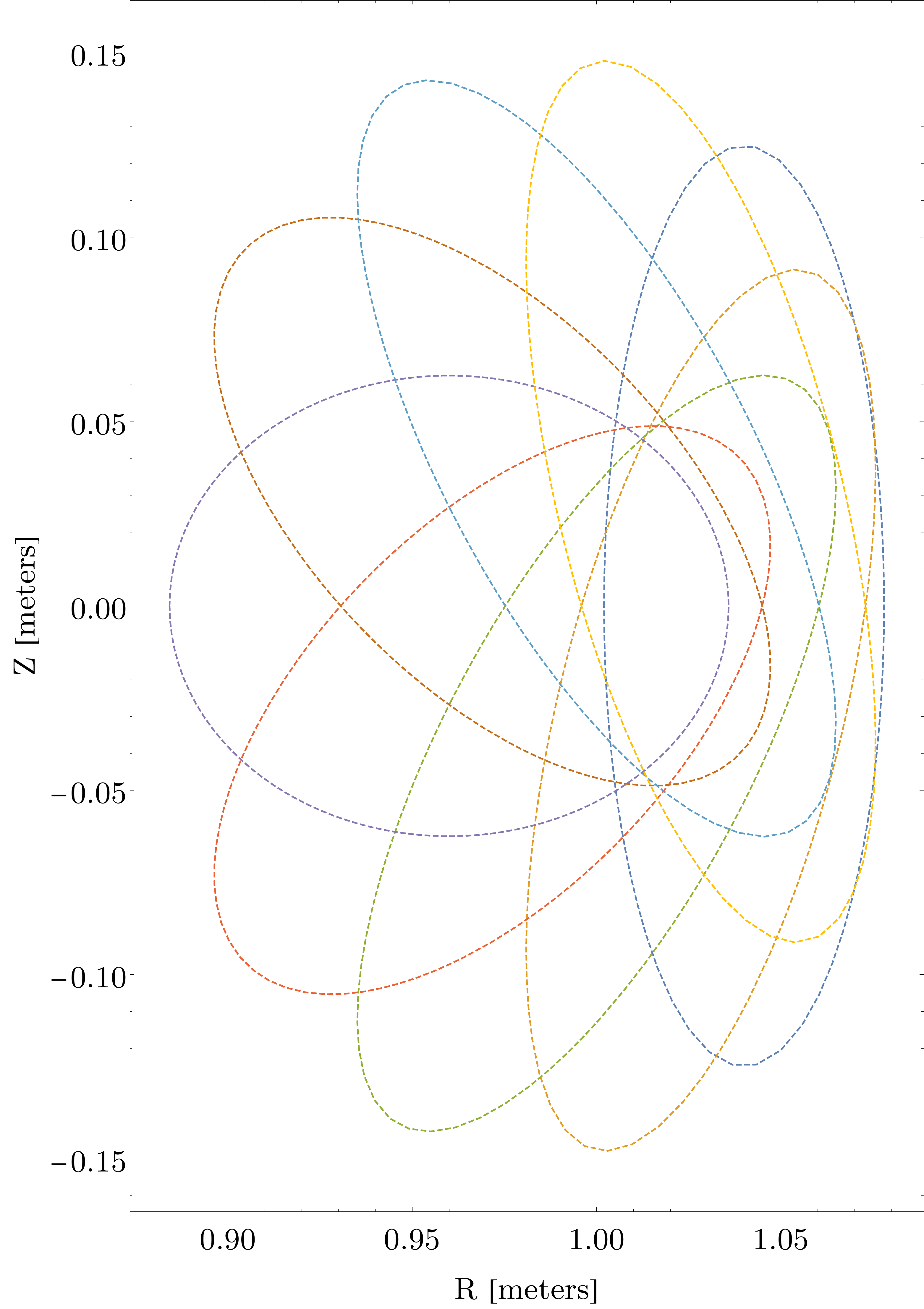}
    \caption{Left: Plasma boundary surface at $\psi=0.015$ Tm\textsuperscript{2} and $B\textsubscript{0}=1$ T of the first order quasisymmetric solution using the axis in \cref{eq:r0qs2sol1} and $\overline \eta=0.7$. The colors show the strength of the magnetic field. Right: Eight poloidal planes within one field period of the corresponding quasisymmetric solution.}
    \label{fig:qs2sol1}
\end{figure}

\begin{figure}
    \centering
    \includegraphics[width=.5\textwidth]{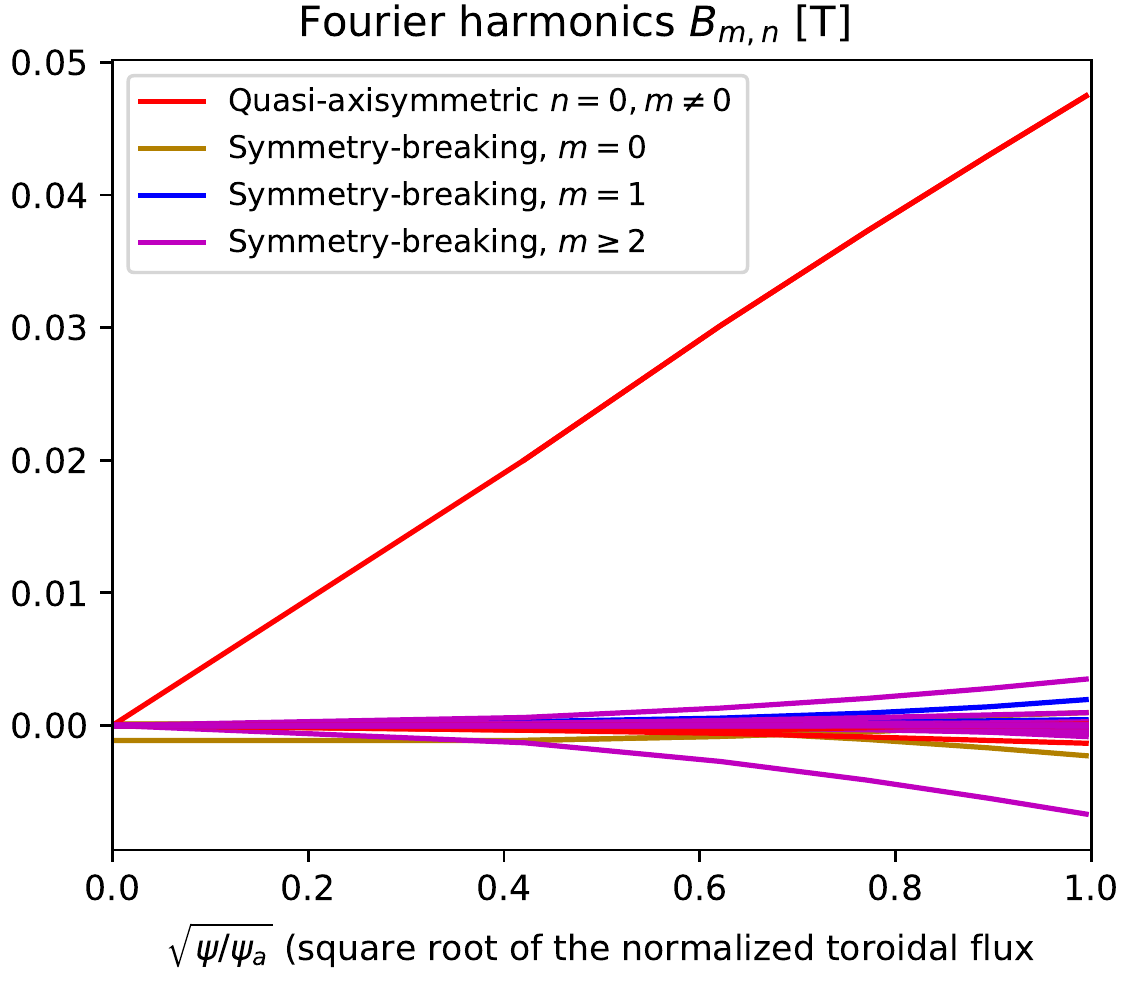}
    \caption{Fourier amplitudes $B\textsubscript{m,n}(\psi)$ of the magnetic field strength computed by BOOZ\_XFORM for the first order quasisymmetric solution using the axis in \cref{eq:r0qs2sol1} and $\overline \eta=0.7$. The horizontal axis displays the square root of the toroidal flux normalized to $\psi\textsubscript{a}=0.015$  Tm\textsuperscript{2}.}
    \label{fig:boozxformqs2sol1}
\end{figure}

As an example of first order quasisymmetry, we choose an axis with three field periods
\begin{equation}
    \mathbf r_0 \text{[m]} = (1+0.04 \cos 3 \Phi_a)\mathbf e_R-0.04 \sin 3 \Phi_a \mathbf e_z.
\label{eq:r0qs2sol1}
\end{equation}
The axis in \cref{eq:r0qs2sol1} has a total of $N=0$ turns of the normal vector after one circuit along the axis yielding a quasi-axisymmetric magnetic field shape.
Choosing $\overline \eta = 0.7$, we obtain for the rotational transform on axis $\iota_0=0.312$.
The parameters $\eta$ and $\delta$ characterizing the surface shape and the corresponding magnetic field are obtained using \cref{eq:qscosheta,eq:qstan2delta}.
Their Fourier coefficients with modulus greater than 0.01 are given by
\begin{equation}
\begin{split}
    \eta&=0.81+0.25 \cos 3 \Phi_a-0.12 \cos 6 \Phi_a+0.06 \cos 9 \Phi_a-0.03 \cos 12 \Phi_a \\
    &+ 0.02 \cos 15 \Phi_a - 0.01 \cos 18 \Phi_a,
\end{split}
\end{equation}
and
\begin{equation}
\begin{split}
    \delta &= -\frac{3}{2}\Phi_a+0.70\sin 3\Phi_a-0.24\sin 6 \Phi_a+0.11 \sin 9 \Phi_a-0.64\sin 12 \Phi_a\\
    &+0.04 \sin 15 \Phi_a-0.03 \sin 18 \Phi_a + 0.02 \sin 21 \Phi_a.
\end{split}
\end{equation}
The resulting boundary shape and poloidal cross-sections at $\psi=0.015$ Tm\textsuperscript{2} and $B\textsubscript{0}=1$ are shown in \cref{fig:qs2sol1}.
The Fourier spectrum of $B$ in Boozer coordinates is computed using the BOOZ\_XFORM code and shown in \cref{fig:boozxformqs2sol1}, where it is seen that the $(M,N)=(1,0)$ harmonic is dominant across all surfaces except at a particular region close to the magnetic axis.
As noted in \cite{Landreman2019b}, the presence of a mirror mode in the axis with $n=0$ may be due to the difference between the ideal quasisymmetric solution for $\psi$ and the computed VMEC equilibrium inside a surface with a small but finite minor radius.

\begin{figure}
    \centering
    \includegraphics[width=.49\textwidth]{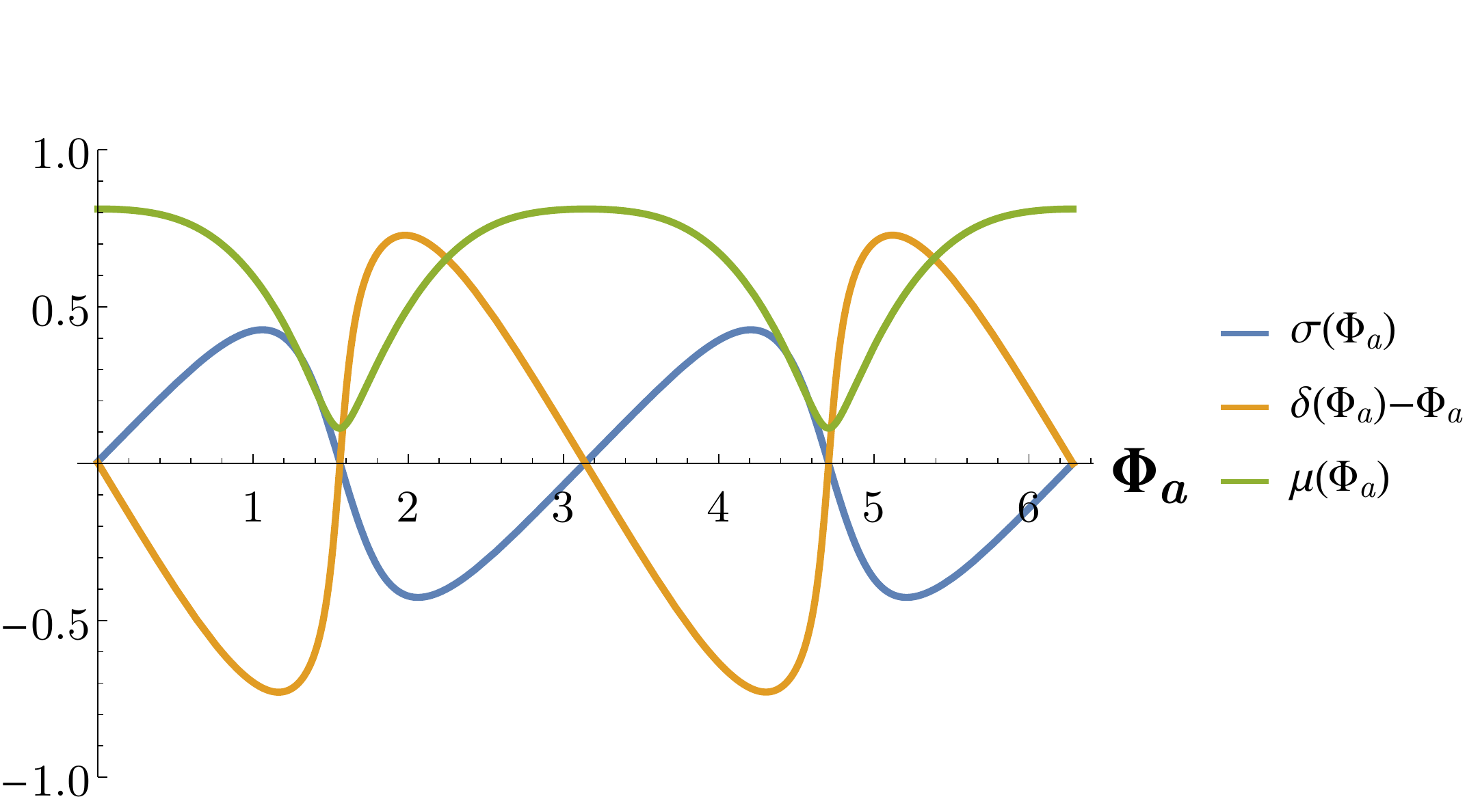}
    \includegraphics[width=.49\textwidth]{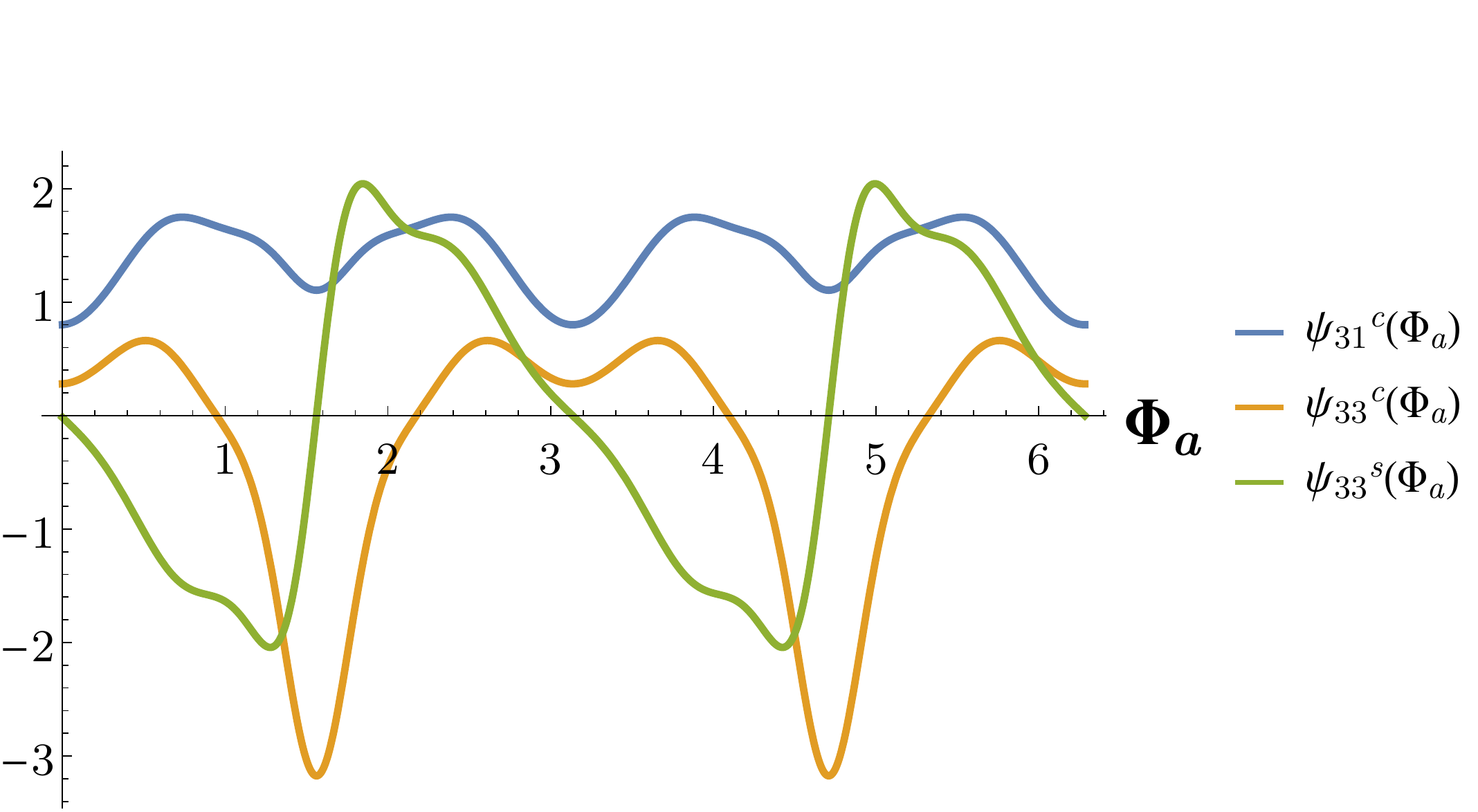}
    \caption{Surface shape parameters of the second order quasi-axisymmetric surface corresponding to the axis in \cref{eq:r0qs3sol2} and $\overline \eta = 0.713$.
    Left: $\sigma, \delta$ and $\mu=\tanh \eta$.
    Right: $\psi\textsubscript{31}\textsuperscript{c}, \psi\textsubscript{33}\textsuperscript{c}$ and $\psi\textsubscript{33}\textsuperscript{s}$.}
    \label{fig:qs3sol21}
\end{figure}

\begin{figure}
    \centering
    \includegraphics[width=.64\textwidth]{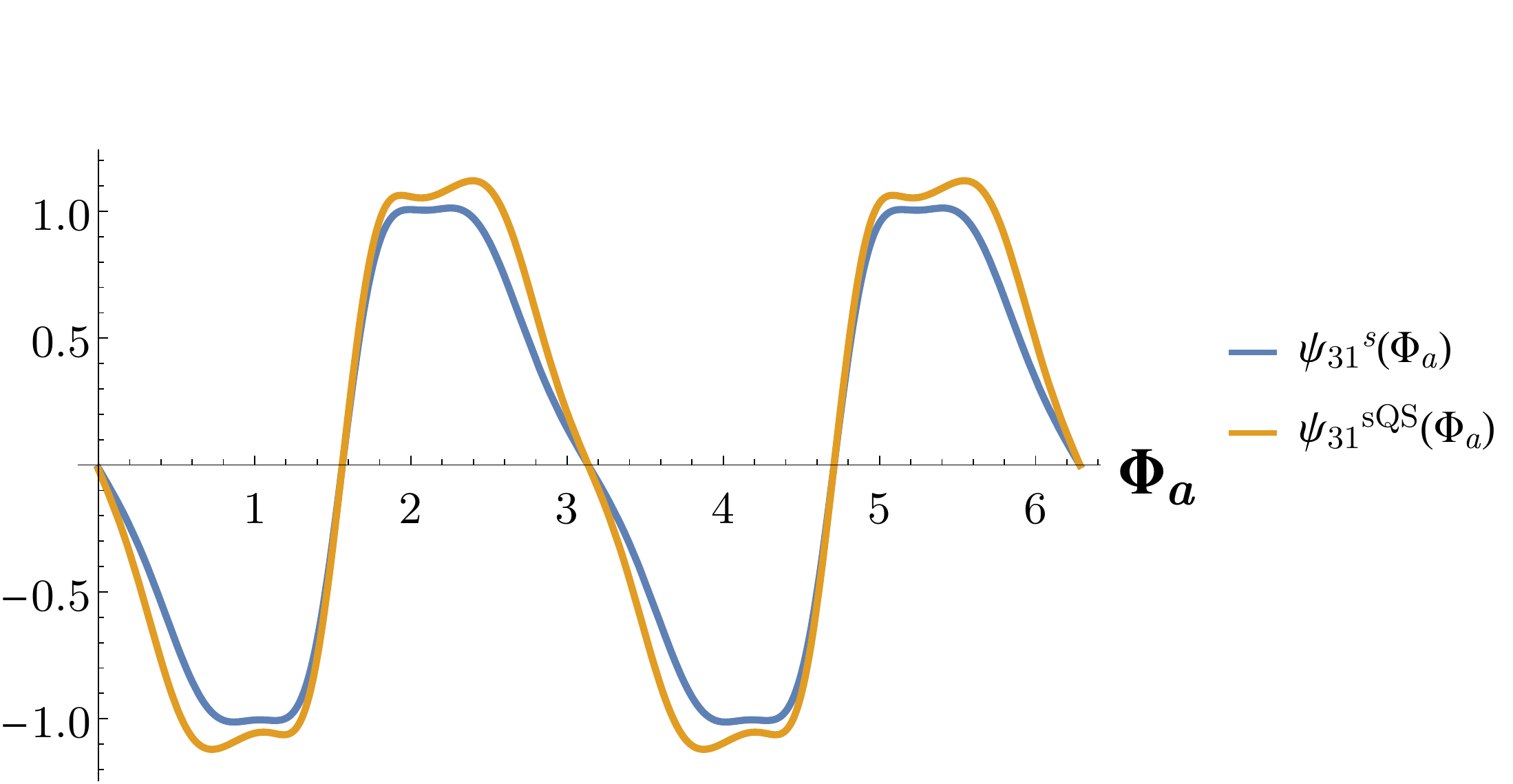}
    \includegraphics[width=.35\textwidth]{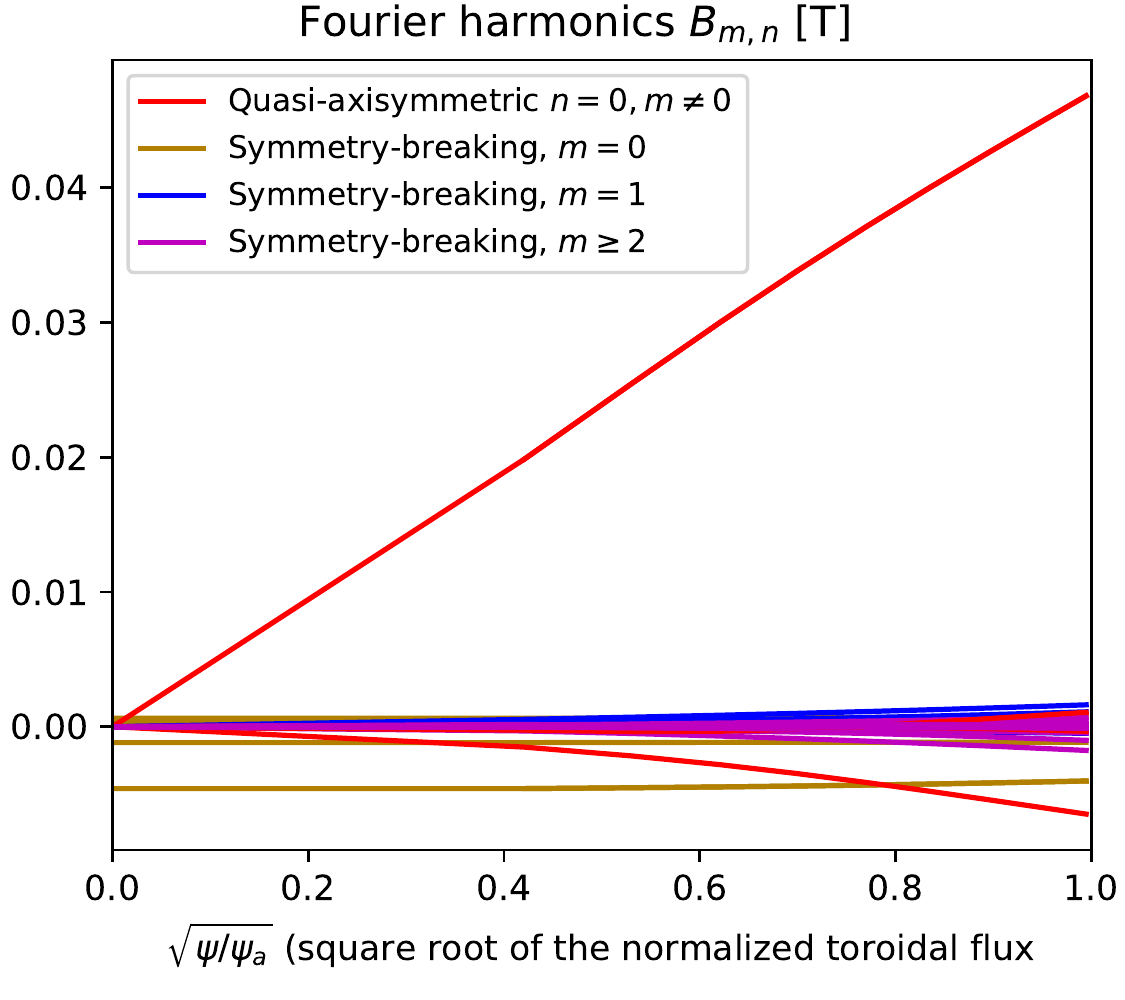}
    \caption{Left: Surface shape parameters $\psi\textsubscript{31}\textsuperscript{s}$ and the quasisymmetry constraint $\psi\textsubscript{31}\textsuperscript{sQS}$ of the second order quasi-axisymmetric surface corresponding to the axis in \cref{eq:r0qs3sol2} and $\overline \eta = 0.713$. Right: Fourier amplitudes $B\textsubscript{m,n}(\psi)$ of the magnetic field strength computed by BOOZ\_XFORM.}
    \label{fig:qs3sol22}
\end{figure}

\begin{figure}
    \centering
    \includegraphics[trim=100 0 60 110,clip,width=.58\textwidth]{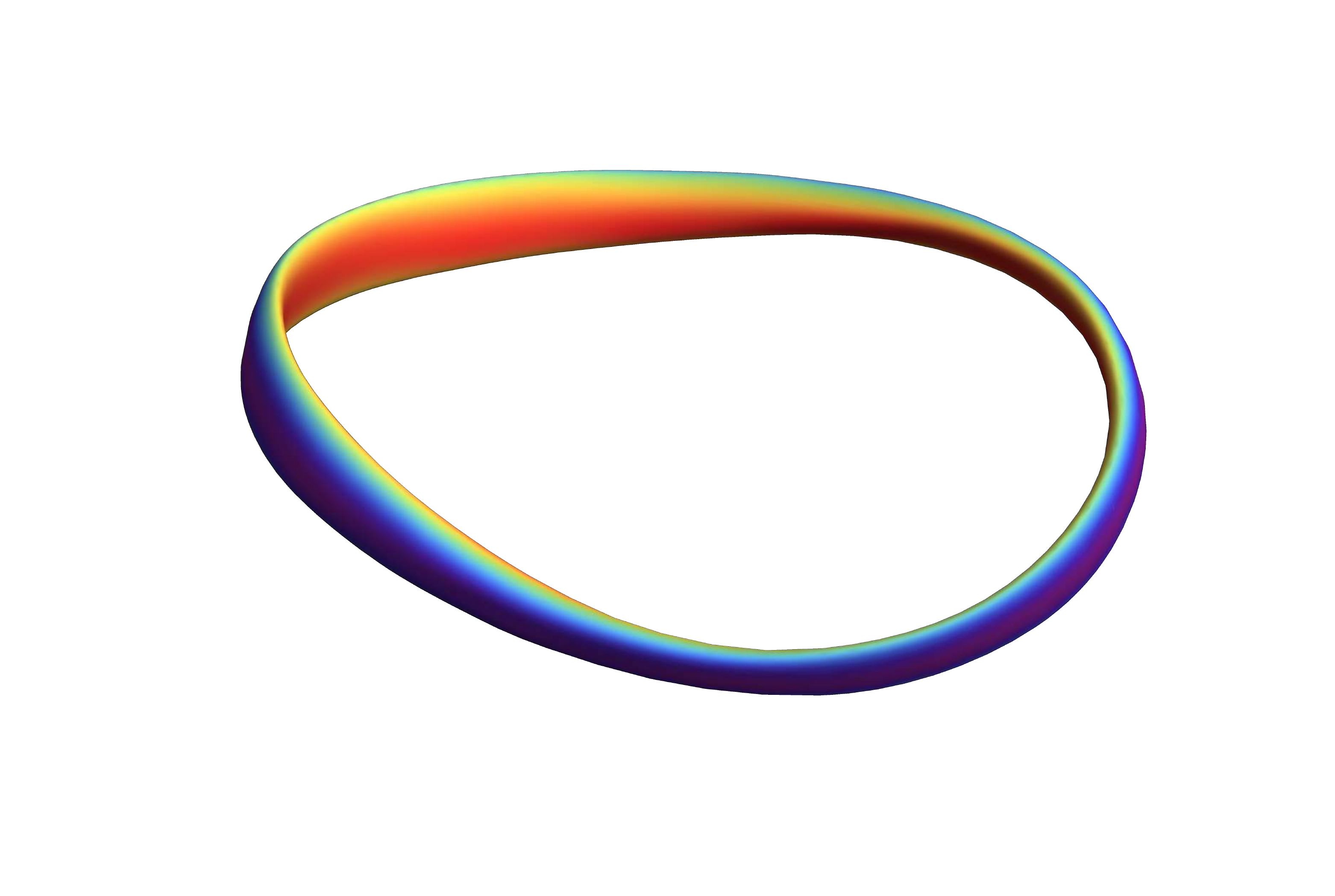}
    \includegraphics[width=.41\textwidth]{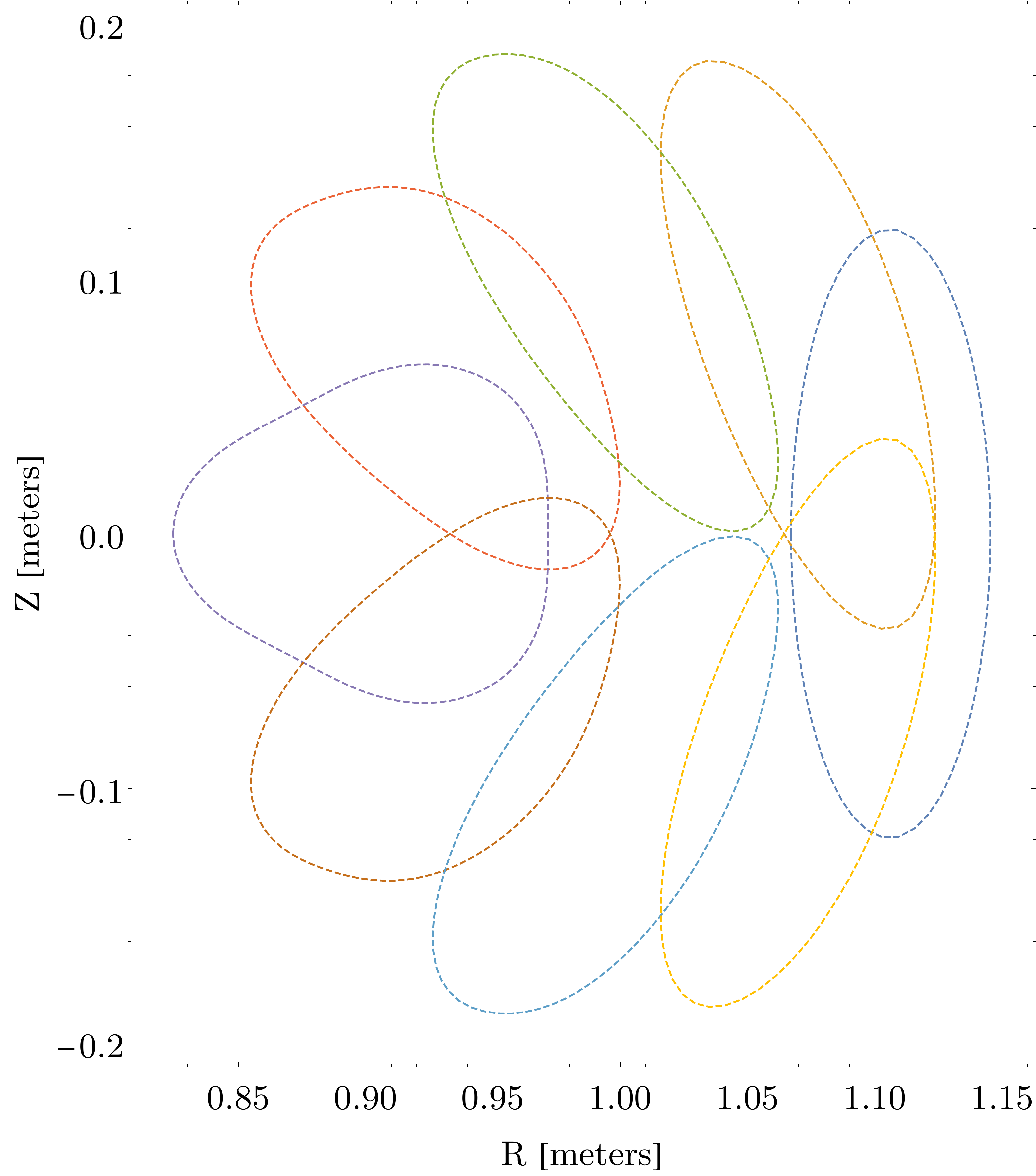}
    \caption{Left: Plasma boundary surface at $\psi=0.015$ Tm\textsuperscript{2} and $B\textsubscript{0}=1$ T of the second order quasisymmetric solution using the axis in \cref{eq:r0qs3sol2} and $\overline \eta=0.713$. The colors show the strength of the magnetic field. Right: Eight poloidal planes within one field period of the corresponding quasisymmetric solution.}
    \label{fig:qs3sol23}
\end{figure}

For second order quasisymmetry, we consider a quasi-axisymmetric solution with two field periods.
The considered axis shape is given by
\begin{align}
    \mathbf r_0 \text{[m]} &= (1+0.101\cos 2 \Phi_a+0.005 \cos 4 \Phi_a)\mathbf e_R+(0.101 \sin 2 \Phi_a+ 0.006 \sin 4 \Phi_a) \mathbf e_z,
\label{eq:r0qs3sol2}
\end{align}
together with $\overline \eta = 0.713$.
The rotational transform on axis is then given by $\iota_0=0.181$.
The shape parameters $\eta, \delta$ and $\sigma$ are shown in \cref{fig:qs3sol21} (left), while their Fourier coefficients with modulus greater than 0.01 are found to be
\begin{equation}
\begin{split}
    \eta &= 0.71+0.29 \cos 2 \Phi_a-0.11 \cos 4 \Phi_a+0.04 \cos 6 \Phi_a-0.02 \cos 8 \Phi_a\\
    &+0.01 \cos 10 \Phi_a-0.01 \cos 12 \Phi_a+0.01 \cos 14 \Phi_a,
\end{split}
\end{equation}
and
\begin{equation}
\begin{split}
    \delta &= \Phi_a-0.68\sin 2\Phi_a+0.23\sin 4 \Phi_a-0.11 \sin 6 \Phi_a+0.06\sin 8 \Phi_a\\
    &-0.04\sin 10 \Phi_a+0.03\sin 12 \Phi_a-0.02\sin 14 \Phi_a.
\end{split}
\end{equation}
The higher order parameters $\psi_{31}^c, \psi_{31}^s, \psi_{33}^c$ and $\psi_{33}^s$ are shown in \cref{fig:qs3sol21} (right) and \cref{fig:qs3sol22}.
The resulting boundary shape and poloidal cross-sections at $\psi=0.015$ Tm\textsuperscript{2} and $B\textsubscript{0}=1$ are shown in \cref{fig:qs3sol23}.
The departure from quasisymmetry is evaluated in \cref{fig:qs3sol22} by comparing the obtained $\psi_{31}^s$ function with the quasisymmetry constraint $\psi_{31}^{sQS}$ given by \cref{eq:sandra1} where, in general, a good agreement is found, and with the results from the BOOZ\_XFORM code.
A mirror mode with a Fourier harmonic $m=0$ of $B$ on the axis is also present in this configuration albeit with larger amplitude relative to the quasi-axisymmetric mode when compared with the mirror mode present the first order configuration.
As noted in Ref. \cite{Landreman2019b} such worsening of the mirror mode on the axis is expected as the difference between the ideal near-axis solution and the VMEC solution inside a finite minor radius is worsened at higher order in the expansion. 

We remark that the quasisymmetric solutions presented in this section are not exhaustive.
Their purpose is merely to illustrate how near-axis quasisymmetric solutions might be obtained using the direct coordinate approach.
Although the examples shown here resemble previously obtained quasisymmetric solutions using other approaches \cite{Landreman2019a,Landreman2019b}, a more thorough comparison with well known numerical tools such as VMEC and BOOZ\_XFORM, together with a comparison with numerical tools solving quasisymmetry with the inverse coordinate approach is needed in order to establish a definite numerical tool ready to be used in the design of realistic experiments.

\section{Conclusions}
\label{sec:conc}

In this work, a direct coordinate approach is used to construct quasisymmetric magnetic fields up to second order in the distance to the magnetic axis.
The direct coordinate approach allows us to derive the constraints for quasisymmetric magnetic fields using an orthogonal coordinate system tied to the magnetic axis.
Our major findings consist on showing that quasisymmetry is broken at the third order in the expansion parameter $\epsilon$ and that quasisymmetry might be achievable on a single flux surface using a formulation independent of the use of Boozer coordinates.
In addition, we were able to provide different formulations of quasisymmetry in terms of Mercier coordinates at different orders in the expansion, shedding some light on the physics of quasisymmetric magnetic field configurations.
In particular, we are able to link the parameters and constraints of first order direct coordinate approach with the inverse one and derive the direction of the quasisymmetry vector in both formulations.
A similar comparison for second order quasisymmetry is left for future work.
Finally, using the analytical expressions for the quasisymmetry constraints, we were able to numerically generate quasi-axisymmetric designs both at first and second order in the expansion parameter.

As an alternative avenue of future study, we mention the possibility of improving the methods developed here to generate second order quasisymmetric shapes by obtaining the coordinate transformation between Mercier and Boozer coordinates at next order in the expansion, generalizing the magnetic field strength derived in \cref{eq:magfieldboozmerc}.
This would allow us to directly assess the constants present in the second order system of equations.
A thorough assessment of the accuracy of the second order numerical solutions using the VMEC and BOOZER\_XFORM codes, which requires the double expansion formulation developed in Ref. \cite{Landreman2019b}, is also left for future work.
We also mention the possibility of obtaining exact second order quasisymmetric configurations by only partially specifying the axis, e.g. letting the curvature or torsion be an output parameter of the system.

\section{Acknowledgements}

This work was supported by a grant from the Simons Foundation (560651, ML) and a US DOE Grant No. DEFG02-86ER53223.

\appendix

\section{Second Order Quasisymmetry Equations}

In this appendix, we present the four rows of the $M_4$ matrix of \cref{eq:qseqs2fin}, namely $M_{41}, M_{42}, M_{43}$ and $M_{44}$, needed to obtain the third order quasisymmetric toroidal flux $\psi$.
The four rows of $M_4=\overline M_4/({8 F_0 {\overline \eta}^4 \kappa^8})$ are then given by 
\begin{dmath}
\overline M_{41}=64 \pi ^3 B_0^3 F_0^3 {\overline \eta}^{10} \kappa'-4 {\overline \eta}^4 \kappa^5 \left(2 \sigma \left(4 \pi ^3 B_0^3 F_0^3 {\overline \eta}^2 \sigma'+7 \pi  B_0 F_0 \kappa^{'2} \tau\right)+4 \pi  B_0 F_0 {\overline \eta}^2 \left(\pi  B_0 F_0 \sigma''+4 \tau \tau'\right)+5 \kappa' \kappa'' \tau\right)+2 {\overline \eta}^2 \kappa^7 \left(-16 \pi ^2 B_0^2 F_0^2 {\overline \eta}^2 \sigma^2 \tau'+2 \pi  B_0 F_0 \sigma \left({\overline \eta}^2 \left(\pi  B_0 F_0 \left({\overline \eta}^2-16 \tau \sigma'\right)+2 \tau''\right)-5 \kappa^{'2} \sigma'\right)-2 {\overline \eta}^2 \tau \left(2 \pi  B_0 F_0 \sigma''+5 \tau \tau'\right)-5 \kappa' \kappa'' \sigma'\right)+16 \pi  B_0 F_0 {\overline \eta}^6 \kappa^2 \kappa' \left(10 \pi  B_0 F_0 {\overline \eta}^2 \tau+3 \kappa^{'2}\right)-8 \pi  B_0 F_0 {\overline \eta}^6 \kappa^3 \left(-2 \pi  B_0 F_0 \sigma \kappa^{'2}+6 \pi  B_0 F_0 {\overline \eta}^2 \tau'+5 \kappa' \kappa''\right)+2 {\overline \eta}^2 \kappa^8 \left(\kappa' \left(2 \pi  B_0 F_0 \sigma' \left(4 \pi  B_0 F_0 \left(2 \sigma^2-1\right)-3 \sigma'\right)+3 {\overline \eta}^2 \tau+10 \tau^2 \sigma'\right)+\sigma' \left(\kappa'''-2 \pi  B_0 F_0 \sigma \kappa''\right)\right)-8 \kappa^4 \left(4 \pi  B_0 F_0 {\overline \eta}^6 \kappa' \left(\pi  B_0 F_0 \left(2 \pi  B_0 F_0+\sigma'\right)-3 \tau^2\right)+\pi  B_0 F_0 {\overline \eta}^6 \left(2 \pi  B_0 F_0 \sigma \kappa''-\kappa'''\right)-3 {\overline \eta}^4 \kappa^{'3} \tau\right)+4 {\overline \eta}^2 \kappa^6 \left(2 {\overline \eta}^2 \kappa' \left(\pi  B_0 F_0 \left({\overline \eta}^2+2 \sigma \tau'\right)+4 \pi  B_0 F_0 \tau \left(\sigma'-\pi  B_0 F_0\right)+2 \tau^3\right)+{\overline \eta}^2 \tau \left(6 \pi  B_0 F_0 \sigma \kappa''+\kappa'''\right)+3 \kappa^{'3} \sigma'\right)-3 \kappa^{11} \sigma' \left(\pi  B_0 F_0 \sigma \left(5 {\overline \eta}^2+4 \tau \sigma'\right)-\tau' \sigma'\right)-2 {\overline \eta}^2 \kappa^9 \left(\pi  B_0 F_0 \sigma \left(-4 \pi  B_0 F_0 \sigma' \left(4 \pi  B_0 F_0+3 \sigma'\right)+9 {\overline \eta}^2 \tau+8 \tau^2 \sigma'\right)+2 \sigma' \left(\pi  B_0 F_0 \sigma''+\tau \tau'\right)\right)+3 \kappa^{10} \kappa' \sigma' \left({\overline \eta}^2+2 \tau \sigma'\right),
\label{eq:bb1}
\end{dmath}
\begin{dmath}
\overline  M_{42}=2 \kappa \left(2 \kappa^6 \left(4 \pi ^2 B_0^2 F_0^2 {\overline \eta}^4 \left(2 \sigma' \left(\pi  B_0 F_0 \left(\sigma^2+1\right)+\sigma'\right)+\sigma \sigma''\right)+\tau \left({\overline \eta}^4 \left(\tau''-\pi  B_0 F_0 \left({\overline \eta}^2-6 \sigma \tau'\right)\right)-{\overline \eta}^2 \kappa^{'2} \sigma'\right)+2 \pi  B_0 F_0 {\overline \eta}^4 \tau^2 \sigma'\right)+48 \pi ^2 B_0^2 F_0^2 {\overline \eta}^8 \kappa^{'2}+2 {\overline \eta}^4 \kappa^4 \left(20 \pi ^2 B_0^2 F_0^2 \sigma^2 \kappa^{'2}-2 \kappa^{'2} \left(\tau^2-3 \pi  B_0 F_0 \sigma'\right)+2 \pi  B_0 F_0 \sigma \left(2 \pi  B_0 F_0 {\overline \eta}^2 \tau'+5 \kappa' \kappa''\right)+\pi  B_0 F_0 {\overline \eta}^2 \left(2 \tau''-3 \pi  B_0 F_0 {\overline \eta}^2\right)\right)-16 \pi ^2 B_0^2 F_0^2 {\overline \eta}^8 \kappa \left(2 \pi  B_0 F_0 \sigma \kappa'+\kappa''\right)+4 {\overline \eta}^4 \kappa^5 \left(-2 \pi ^2 B_0^2 F_0^2 \sigma^2 \kappa''+\kappa'' \left(\tau^2-2 \pi  B_0 F_0 \sigma'\right)+\pi  B_0 F_0 \sigma \left(2 \kappa' \left(\pi  B_0 F_0 \left(4 \pi  B_0 F_0+5 \sigma'\right)-2 \tau^2\right)-\kappa'''\right)+\kappa' \tau \tau'\right)+{\overline \eta}^2 \kappa^8 \left(-4 \pi ^2 B_0^2 F_0^2 {\overline \eta}^2-8 \pi ^2 B_0^2 F_0^2 \sigma^2 \left({\overline \eta}^2-3 \tau \sigma'\right)-10 \pi  B_0 F_0 \sigma \tau' \sigma'+\sigma' \left(8 \pi  B_0 F_0 \tau \left(\pi  B_0 F_0+\sigma'\right)+\tau''+2 \tau^3\right)\right)+16 \pi  B_0 F_0 {\overline \eta}^6 \kappa^2 \kappa^{'2} \tau+2 {\overline \eta}^2 \kappa^7 \kappa' \left(2 \pi  B_0 F_0 \sigma \left({\overline \eta}^2-\tau \sigma'\right)+\tau' \sigma'\right)+8 \pi  B_0 F_0 {\overline \eta}^4 \kappa^3 \kappa' \left({\overline \eta}^2 \tau'-\sigma \left(8 \pi  B_0 F_0 {\overline \eta}^2 \tau+3 \kappa^{'2}\right)\right)+\kappa^9 \sigma^{'2} \left(4 \pi  B_0 F_0 \sigma \kappa'-\kappa''\right)+\kappa^{10} \sigma^{'2} \left(4 \pi  B_0 F_0 \left(\pi  B_0 F_0+\sigma'\right)+\tau^2\right)\right),
 \label{eq:bb2}
\end{dmath}
\begin{dmath}
\overline  M_{43}= \kappa^2 \left(32 \pi ^2 B_0^2 F_0^2 {\overline \eta}^8 \kappa' \tau+8 \kappa^2 \left(2 \pi  B_0 F_0 {\overline \eta}^6 \kappa' \left(\pi  B_0 F_0 \sigma'+4 \tau^2\right)+3 {\overline \eta}^4 \kappa^{'3} \tau\right)+2 {\overline \eta}^2 \kappa^5 \left(2 \pi  B_0 F_0 \sigma \left(2 \pi  B_0 F_0 {\overline \eta}^2 \left({\overline \eta}^2-2 \tau \sigma'\right)-5 \kappa^{'2} \sigma'\right)-2 {\overline \eta}^2 \left(2 \pi  B_0 F_0 \tau \sigma''+\tau' \left(3 \pi  B_0 F_0 \sigma'+5 \tau^2\right)\right)-5 \kappa' \kappa'' \sigma'\right)-4 {\overline \eta}^4 \kappa^3 \tau \left(5 \kappa' \left(2 \pi  B_0 F_0 \sigma \kappa'+\kappa''\right)+6 \pi  B_0 F_0 {\overline \eta}^2 \tau'\right)-2 {\overline \eta}^2 \kappa^6 \left(\kappa' \left(2 \pi  B_0 F_0 \sigma' \left(4 \pi  B_0 F_0+3 \sigma'\right)+{\overline \eta}^2 \tau-10 \tau^2 \sigma'\right)-\sigma' \left(2 \pi  B_0 F_0 \sigma \kappa''+\kappa'''\right)\right)+4 {\overline \eta}^2 \kappa^4 \left({\overline \eta}^2 \kappa' \left(-\pi  B_0 F_0 {\overline \eta}^2+2 \pi  B_0 F_0 \tau \left(\sigma'-4 \pi  B_0 F_0\right)+4 \tau^3\right)+{\overline \eta}^2 \tau \left(2 \pi  B_0 F_0 \sigma \kappa''+\kappa'''\right)+3 \kappa^{'3} \sigma'\right)-3 \kappa^9 \sigma' \left(\pi  B_0 F_0 \sigma \left({\overline \eta}^2+4 \tau \sigma'\right)-\tau' \sigma'\right)-2 {\overline \eta}^2 \kappa^7 \left(\pi  B_0 F_0 \sigma \left(4 \pi  B_0 F_0 \sigma^{'2}+{\overline \eta}^2 \tau+12 \tau^2 \sigma'\right)+2 \sigma' \left(\pi  B_0 F_0 \sigma''+\tau \tau'\right)\right)+\kappa^8 \kappa' \sigma' \left(6 \tau \sigma'-{\overline \eta}^2\right)\right),
\label{eq:bb3}
\end{dmath}
\begin{dmath}
\overline  M_{44}=2 \kappa^3 \left(-4 {\overline \eta}^4 \kappa^2 \left(\pi ^2 B_0^2 F_0^2 {\overline \eta}^4+\kappa^{'2} \left(\tau^2-3 \pi  B_0 F_0 \sigma'\right)\right)+{\overline \eta}^2 \kappa^6 \left(\sigma' \left(\pi  B_0 F_0 \left({\overline \eta}^2-4 \sigma \tau'\right)+8 \pi  B_0 F_0 \tau \left(\pi  B_0 F_0+\sigma'\right)+\tau''+2 \tau^3\right)-2 \pi ^2 B_0^2 F_0^2 {\overline \eta}^2\right)-8 \pi  B_0 F_0 {\overline \eta}^6 \kappa \kappa'' \tau+24 \pi  B_0 F_0 {\overline \eta}^6 \kappa^{'2} \tau+2 {\overline \eta}^2 \kappa^5 \kappa' \left(\tau' \sigma'-\pi  B_0 F_0 \sigma \left({\overline \eta}^2-4 \tau \sigma'\right)\right)-2 {\overline \eta}^2 \kappa^4 \tau \left({\overline \eta}^2 \left(4 \pi  B_0 F_0 \sigma \tau'-\tau''\right)+\kappa^{'2} \sigma'\right)+4 {\overline \eta}^4 \kappa^3 \left(\kappa'' \left(\tau^2-\pi  B_0 F_0 \sigma'\right)+\kappa' \tau \tau'\right)+\kappa^7 \sigma^{'2} \left(4 \pi  B_0 F_0 \sigma \kappa'-\kappa''\right)+\kappa^8 \sigma^{'2} \left(4 \pi  B_0 F_0 \left(\pi  B_0 F_0+\sigma'\right)+\tau^2\right)\right).
\label{eq:bb4}
\end{dmath}

\section{Quasisymmetry Condition}
\label{app:qscond}

Here we will show that quasisymmetry is sufficient for omnigenity.
The Lagrangian $\mathcal L$ for guiding-centre motion can be derived by expanding the Lagrangian of a charged particle immersed in an electromagnetic field up to first order in $\rho^*=\rho_{th}/a$, with $\rho_{th}=v_{th}/\Omega$ the thermal Larmor radius, $v_{th}=\sqrt{T/m}$ the thermal velocity, $\Omega=q B/m$ the gyrofrequency, $T$ the temperature of the particle species, $m$ its mass and $q$ its charge, yielding \cite{Littlejohn1983a,Cary2009,Jorge2017,Frei2019}
\begin{equation}
    \mathcal L = (q \mathbf A + m v_\parallel \mathbf b) \cdot \dot{\mathbf r}+\frac{\mu \dot \Theta}{m}-\frac{m v_\parallel^2}{2}-\mu B.
\label{eq:gclag}
\end{equation}
In \cref{eq:gclag} we defined $\mathbf{A}$ as the magnetic vector potential, $\mathbf r = \mathbf x - \bm \rho$ the particle guiding-center, $\mathbf x$ the particle position, $\bm \rho=\mathbf v \times \mathbf b/\Omega$ the particle Larmor radius, $\mathbf v$ the particle velocity, $v_\parallel = \mathbf v \cdot \mathbf b$ the particle parallel velocity, $\mu = m v_\perp^2/(2B)$ the magnetic moment and $\Theta$ the gyroangle.
We note that, in \cref{eq:gclag}, the spatial dependence of $\mathcal{L}$ is introduced only via $\mathbf A=\mathbf A(\mathbf r)$, $\mathbf b = \mathbf B/B = \mathbf b(\mathbf r)$ and $B = B(\mathbf r)$, and electric fields are neglected for simplicity.
The equations of motion for the phase-space coordinates $(\mathbf r, v_\parallel, \mu, \theta)$ can be found by applying the Euler-Lagrange equations to the Lagrangian in \cref{eq:gclag}, yielding the guiding center velocity
\begin{equation}
    \dot{\mathbf r}=v_\parallel \mathbf b+\frac{v_\parallel^2}{\Omega^*}\mathbf b \times \mathbf k+\frac{\mu B}{m \Omega^*}\frac{\mathbf b \times \nabla B}{B},
\label{eq:dotr}
\end{equation}
the parallel acceleration
\begin{equation}
    \dot v_\parallel =\Omega \frac{\mu B}{m \Omega^*}\frac{\mathbf B^* \cdot \nabla B}{B^2} ,
\label{eq:dotv}
\end{equation}
the gyrofrequency $\dot \Theta = \Omega = qB/m$ and $\dot \mu=0$.
In \cref{eq:dotr,eq:dotv}, we defined $\mathbf B^*=B(\mathbf b + v_\parallel \nabla \times \mathbf b/\Omega)$ and $\Omega^*=q B_\parallel^*/m$ with $B_\parallel^*=\mathbf B^*\cdot \mathbf b$.
We note that, according to \cref{eq:dotr,eq:dotv}, the energy $H=mv_\parallel^2/2+\mu B$ is a constant of motion, as $\dot H = \partial H/\partial t + \dot{\mathbf r}\cdot \nabla H+\dot v_\parallel \partial H/\partial v_\parallel=0$.
The parallel velocity $v_\parallel$ can then be expressed as a function of the magnetic field strength as
\begin{equation}
    v_\parallel = \pm \sqrt{\frac{2(H-\mu B)}{m}},
\label{eq:vpar}
\end{equation}

For good confinement, we would like to have omnigenity, which is the condition that the net radial displacement $\Delta \psi = 0$ for all values of $\mu$.
Therefore, we calculate the net radial displacement $\Delta \psi$ of trapped particles in an arbitrary magnetic configuration using the equations of motion in \cref{eq:dotr,eq:dotv}.
The quantity $\Delta \psi$ is derived by integrating the temporal derivative $\dot \psi$ over the time it takes for the particle to go from one bouncing point to another along a particular magnetic field line
\begin{equation}
    \Delta \psi = \int \dot \psi dt = \int_{l_0}^{l_1} \left(\frac{\partial \psi}{\partial t}+\dot{\mathbf r}\cdot \nabla \psi+\dot v_\parallel \frac{\partial \psi}{\partial v_\parallel}\right)\frac{dl}{v_\parallel},
\label{eq:deltapsi1}
\end{equation}
where $l$ is the arc length along the field line, i.e. $\partial \mathbf x/\partial l = \mathbf b$ and $\partial \psi/\partial l = \mathbf b \cdot \nabla \psi = 0$, and the last integral in \cref{eq:deltapsi1} is performed between two bouncing points $l_0$ and $l_1$.
Noting that for static fields $\partial \psi/\partial t =\partial \psi/\partial v_\parallel = 0$, and using the expression for $\dot{\mathbf r}$ in \cref{eq:dotr}, we find
\begin{equation}
    \Delta \psi = \int_{l_0}^{l_1} \frac{\mathbf B \times \nabla \psi \cdot \nabla B}{\mathbf B \cdot \nabla B}\left.\frac{\partial\left({v_\parallel}/{B}\right)}{\partial l}\right|_{H,\mu}\frac{dl}{1+\frac{v_\parallel \mathbf b \cdot \nabla \times \mathbf b}{\Omega}},
\label{eq:deltapsi}
\end{equation}
where the partial derivative in \cref{eq:deltapsi} is taken at constant $H$ and $\mu$ using the expression for $v_\parallel$ in \cref{eq:vpar}.
In order to make $\Delta \psi=0$, we perform the choice
\begin{equation}
    \frac{\mathbf B \times \nabla \psi \cdot \nabla B}{\mathbf B \cdot \nabla B}=\frac{1}{F(\psi)},
\label{eq:qscond11}
\end{equation}
with $F=F(\psi)$ a flux a function.
Equation (\ref{eq:qscond11}) is the quasisymmetry condition.
By focusing on vacuum fields where ${\mathbf b \cdot \nabla \times \mathbf b}=0$, together with the quasisymmetry condition in \cref{eq:qscond11}, the integral in \cref{eq:deltapsi} yields
\begin{equation}
    \Delta \psi = F(\psi) \left(\left.\frac{v_\parallel}{B}\right|_{l_1}-\left.\frac{v_\parallel}{B}\right|_{l_0}\right)=0,
\end{equation}
as, by definition, $v_\parallel$ vanishes on the bouncing points.
For a more detailed discussion on the different forms of the quasisymmetry constraint other than \cref{eq:qscond11} and the proof that \cref{eq:qscond11} is equivalent to the condition that the magnetic field strength depends on a linear combination of poloidal and toroidal Boozer coordinates with integer coefficients $M$ and $N$, see Ref. \cite{Helander2014}.


\section*{References}
\bibliography{library}
\bibliographystyle{unsrt}

\end{document}